%% file: main.tex
\documentclass[twocolumn, 10pt]{article}
\usepackage[a4paper, margin=2.6cm]{geometry}
\usepackage{subcaption}
\usepackage{algorithm}
\usepackage{algorithmic}
\usepackage{tikz}
\usepackage{tkz-euclide}
\usepackage{tikz-3dplot}
\usetikzlibrary{calc}
\usepackage{pgfplots}
\pgfplotsset{compat=1.18}
\usepackage{pgfplotstable}
\usepackage{authblk}
\usepackage{booktabs}
\usepackage{amsmath}
\usepackage{amssymb}
\usepackage{url}
\usepackage{footnote}
\usepackage[pagebackref=true,breaklinks=true,bookmarks=false,hidelinks]{hyperref}

\definecolor{myLightBlue}{RGB}{100,143,255}
\definecolor{myDarkBlue}{RGB}{120,94,240}
\definecolor{myOrange}{RGB}{254,97,0}
\definecolor{myYellow}{RGB}{255,176,0}
\definecolor{mycolorA}{RGB}{255,176,0}
\definecolor{mycolorB}{RGB}{254,97,0}
\definecolor{mycolorC}{RGB}{220,38,127}
\definecolor{mycolorD}{RGB}{120,94,240}
\definecolor{mycolorE}{RGB}{100,143,255}
\colorlet{mycolorYO}{myYellow!30!myOrange}

\begin{document}

\title{Intrinsic Meshing of Closed Surfaces Using Geodesic Distances}

\author[1,2]{Tim Gabriel}
\author[2]{Jean-Fran\c{c}ois Remacle}
\author[1]{Christophe Geuzaine}
\affil[1]{Universit\'e de Li\`ege, Li\`ege, Belgium \\
\texttt{\{tim.gabriel, cgeuzaine\}@uliege.be}}
\affil[2]{Universit\'e Catholique de Louvain, Louvain-la-Neuve, Belgium \\
\texttt{jean-francois.remacle@uclouvain.be}}
\date{}
\maketitle

\begin{figure*}
  \centering
  \includegraphics[width=.5\textwidth]{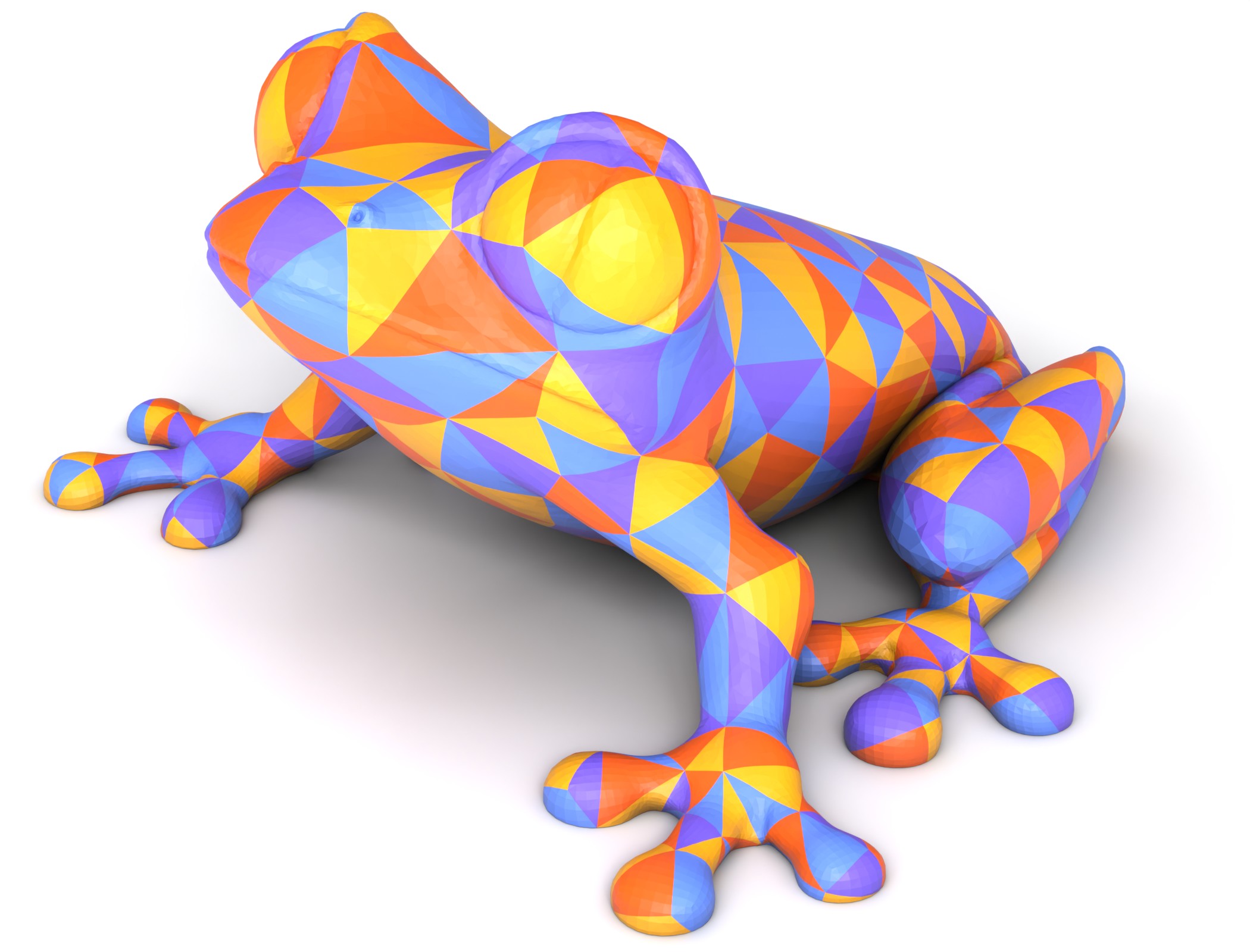}
  \caption{Intrinsic meshing of a surface triangulation from the Thingi10K dataset (ID: 90736), reducing the mesh from 46,000 triangles to 583 intrinsic triangles in under 10 seconds on a standard laptop, using the command: \texttt{gmsh 90736.stl -geodesic -clmin 3 -minIntrinsicAngle 20 -maxIntrinsicAngle 140}.}
  \label{fig:first}
\end{figure*}

\begin{abstract}
We present a method for constructing intrinsic triangulations of closed discrete surfaces, in which edges correspond to shortest geodesic paths and faces decompose into geometric primitives inherited from the underlying mesh. Starting from a watertight input triangulation, the method progressively builds an intrinsic mesh through local optimization operations — edge swaps, edge splits, edge collapses, and triangle splits — performed directly on the surface without modifying the original geometry. Element size is controlled via a characteristic length field, and quality is enforced through angle-based criteria derived from intrinsic distances. Geodesic distances are computed exactly using a continuous Dijkstra approach, accelerated by an A* search strategy that reduces computation to roughly $3\%$ of the cost of standard propagation. The framework supports both refinement and coarsening, overcoming a key limitation of prior intrinsic methods based on developable triangles. As a by-product, the intrinsic triangulation provides a natural foundation for direct high-order mesh generation, bypassing the classical pipeline of first constructing a linear mesh and subsequently curving it. The method is validated on the Thingi10K dataset across nearly 5,000 geometrically complex models.\footnote{The implementation code is openly available within the official Gmsh repository at \url{https://gitlab.onelab.info/gmsh/gmsh}.}
\end{abstract}


\maketitle

% RELATED WORK
\input{relatedWork.tex}

% INTRINSIC OPTERATIONS
\input{intrinsic}

% GEODESIC COMPUTATION
\input{computation}

% RESULT
\input{result}

% FUTURE
\input{future}

\section*{Acknowledgments}
This research was funded in part through the ARC grant for Concerted Research Actions (ARC Discrete-IGA 23/27-08), financed by the Wallonia-Brussels Federation of Belgium. Computational resources have been provided by the Consortium des Équipements de Calcul Intensif (CÉCI), funded by the Fonds de la Recherche Scientifique de Belgique (F.R.S.-FNRS) under Grant No. 2.5020.11 and by the Walloon Region. Human trabecular bone samples were provided by M.A. Hartmann and S. Blouin (Ludwig Boltzmann Institute of Osteology, Vienna). Micro-CT scanning was performed by E. Pedrinazzi (Mechanics of Biological and Bio-Inspired Materials, ULiège).

\bibliographystyle{plain}
\bibliography{references}

\appendix

\input{appendix}

\end{document}

%% file: relatedWork.tex
\section{Introduction}

The numerical representation of surfaces plays a crucial role across a wide range
of domains, including geometric modeling and design, rendering, surface analysis,
geometric processing, and computational physics. Broadly speaking, two main
categories of surface representations can be distinguished: \emph{continuous}
representations and \emph{discrete} representations.

The first category consists of \textbf{continuous surfaces}, typically originating
from design blueprints or analytical descriptions. In computer-aided design (CAD),
surfaces are most commonly defined by explicit parametric mappings
\[
(x,y,z) = \mathbf{x}(u,v),
\]
where $(u,v)$ belong to a parametric domain, usually rectangular. In practice,
these parametric patches are often \emph{trimmed}, meaning that only a subset of
the parameter domain is retained. The trimming curves are themselves typically
defined as parametric curves in the $(u,v)$ space. As a result, CAD models are
generally composed of collections of trimmed parametric patches that must be
assembled to form a complete surface.

Other types of continuous representations also exist. For instance,
\textbf{implicit surfaces} are defined as the zero level set of a scalar function
\[
f(x,y,z) = 0,
\]
as commonly encountered in level-set methods, signed distance functions, or
algebraic surface descriptions.

The second major category corresponds to \textbf{discrete surface representations}.
These arise naturally from numerical approximation, geometric processing, or
measurement procedures such as scanning and reconstruction. Typical examples
include polygonal meshes (most commonly triangle meshes), triangle soups, and
point clouds. Such representations are sometimes referred to as \emph{organic
surfaces}, emphasizing that they do not rely on an underlying analytical
description but rather on sampled geometric data. 

From the perspective of computational physics, our primary objective is to
construct meshes suitable for numerical simulation starting from a wide range
of surface representations, including CAD models, implicit descriptions, or
discrete datasets. These meshes should satisfy two essential requirements:
first, they must exhibit high quality to ensure the accuracy, robustness, and
efficiency of the underlying numerical methods; second, they must remain as
faithful as possible to the original geometry, whether it is defined in a
continuous or discrete manner.

In this work, our objective is to construct \emph{isogeometric} meshes, i.e.
meshes that represent exactly the underlying geometry. When the surface is given
by a parametric description, this objective can in principle be achieved in a
straightforward manner. Consider a surface defined by an injective mapping
\[
\mathbf{x}(u,v) : \Omega \subset \mathbb{R}^2 \rightarrow \mathbb{R}^3 ,
\]
where $\Omega$ is, for instance, a rectangular domain in the parametric space.
A non-overlapping triangulation of $\Omega$ can then be mapped through
$\mathbf{x}(u,v)$ to obtain a triangulation of the surface that represents the
geometry exactly. In other words, an isogeometric triangulation can be obtained
by triangulating the parametric domain and transporting this triangulation to
physical space.

In practice, however, most mesh generation procedures rely on linear triangles
constructed directly in physical space, where only the mesh vertices lie on the
surface. The resulting meshes therefore approximate the geometry rather than
representing it exactly. Nevertheless, when a parametric representation is
available, the construction of isogeometric triangulations remains conceptually
simple.

The situation is significantly more challenging when the surface is defined in a
discrete manner. In this case, no global parametric mapping is available, and the
notion of exact geometric representation must be reconsidered. The purpose of
this paper is precisely to construct isogeometric triangulations for discrete
surfaces, following an approach that is conceptually similar to standard finite
element mesh generation, but adapted to preserve the intrinsic geometry of the
discrete model.

\begin{figure}[b]
  \centering
  \hfill%
  \begin{subfigure}[c]{.4\linewidth}
    \centering
    \includegraphics[width=.9\linewidth]{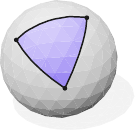}
  \end{subfigure}%
  \hfill%
  \begin{subfigure}[c]{.4\linewidth}
    \centering
    \includegraphics[width=.7\linewidth]{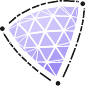}
  \end{subfigure}
  \hfill%
  \hfill%
  \caption{Intrinsic triangle on the sphere (left) and exploded view showing its decomposition into geometric primitives (right).}
  \label{fig:intrinsicTriangle}
\end{figure}

Starting from a watertight triangulation of a closed surface, we
propose to build an \emph{intrinsic triangulation} of that surface, in which the
edges correspond to discrete geodesics and the faces are composed of sets of
small triangles inherited from the initial discrete geometry, as illustrated in Fig.~\ref{fig:intrinsicTriangle}. In this sense, the
new triangulation is defined intrinsically on the surface while remaining fully
consistent with the underlying discrete representation.

Our approach does not rely on modifying the original triangulation. Instead, it
closely follows the philosophy used for meshing CAD surfaces, where a new mesh is
constructed on top of the geometry through local mesh operations. The intrinsic
triangulation is progressively obtained using classical local transformations
such as edge swaps, edge splits, edge collapses, and Delaunay refinement. These
operations are performed intrinsically on the surface, allowing the triangulation
to evolve while preserving the underlying discrete geometry and maintaining a
watertight representation.

We show in this paper that the proposed approach is both robust and sufficiently
fast to be used in a mesh generation context, where users typically expect
algorithms capable of handling large datasets with short turnaround times.
The construction relies only on local operations and discrete geodesic
computations, making it well suited for practical large-scale applications.

As a by-product of this intrinsic construction, we also propose a procedure to
derive from the intrinsic mesh a representation more directly usable in the
context of finite element analysis, namely a \emph{high-order mesh}. The key
feature of this approach is that the high-order geometry is obtained directly
from the intrinsic triangulation, without relying on the classical paradigm in
which a piecewise linear mesh is first generated and subsequently curved.

In that sense, and without much exaggeration, the method can be viewed as a first
attempt to generate high-order meshes without ever passing through the curvature
of a classical piecewise linear mesh. The geometry is instead defined directly
from the intrinsic structure of the surface, with high-order elements naturally
following discrete geodesic edges and intrinsically defined faces.

\section{Related Work}

Simple meshing algorithms have been extensively developed for planar surfaces, 
often based on Delaunay triangulations \cite{ruppert1995delaunay, chew1993guaranteed, shewchuk2002delaunay, gonzaga2012review}. 
These methods aim to represent surfaces using the fewest possible elements while remaining as close as possible 
to the exact geometry and satisfying element quality and size criteria, which may vary spatially. Such planar methods 
can be naturally extended to curved surfaces by computing distances and edges in three-dimensional Euclidean space, 
as demonstrated in \cite{chew1993guaranteed}. In this context, element size criteria can be defined to ensure that the 
mesh accurately captures the surface’s underlying features, including curvature and local geometric 
details \cite{boissonnat2005provably}.

However, although these methods can be used for meshing, they typically rely on a parametrization of 
the original surface, which must be available and may introduce artifacts. In \cite{hoppe1993mesh,wang2001generic}, 
remeshing methods are proposed that rely exclusively on local mesh operations without a parametrization. Nevertheless, 
inserting new points involves a local approximation of the surface, which can introduce inaccuracies even when the input 
surface is exactly known. Regardless of the method, the resulting mesh represents curved geometries using straight-sided 
elements, meaning that accurately capturing such geometries often requires a large number of elements.

\subsection{Intrinsic Meshing}

\begin{figure*}
  \centering
  \hfill%
  \begin{subfigure}{.2\linewidth}
    \centering
    \includegraphics[width=\linewidth]{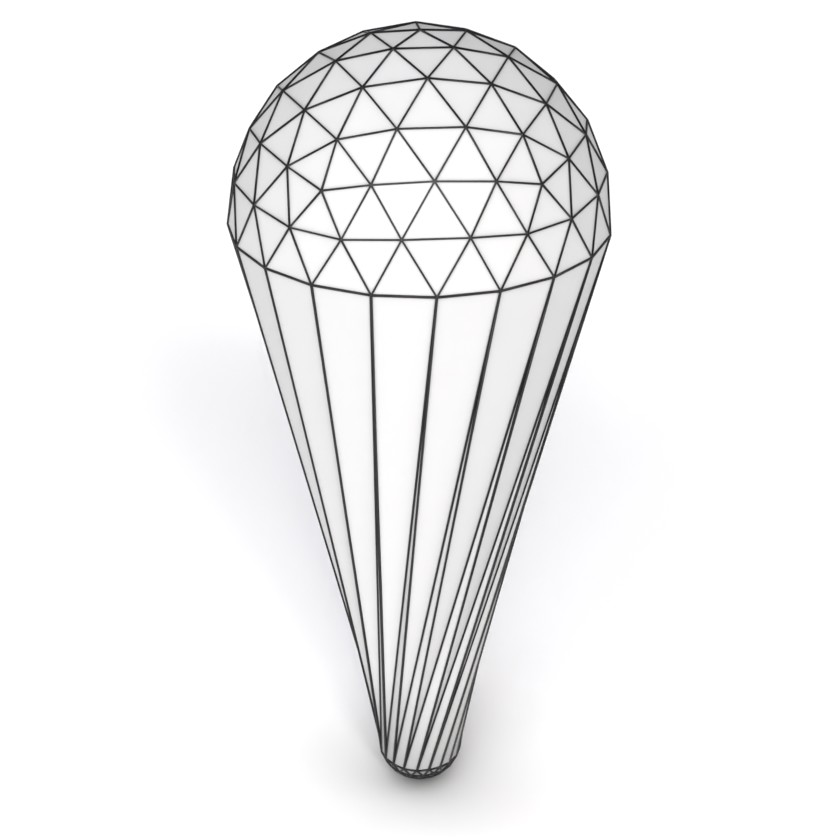}
    \caption{Initial}
  \end{subfigure}%
  \hfill%
  \begin{subfigure}{.2\linewidth}
    \centering
    \includegraphics[width=\linewidth]{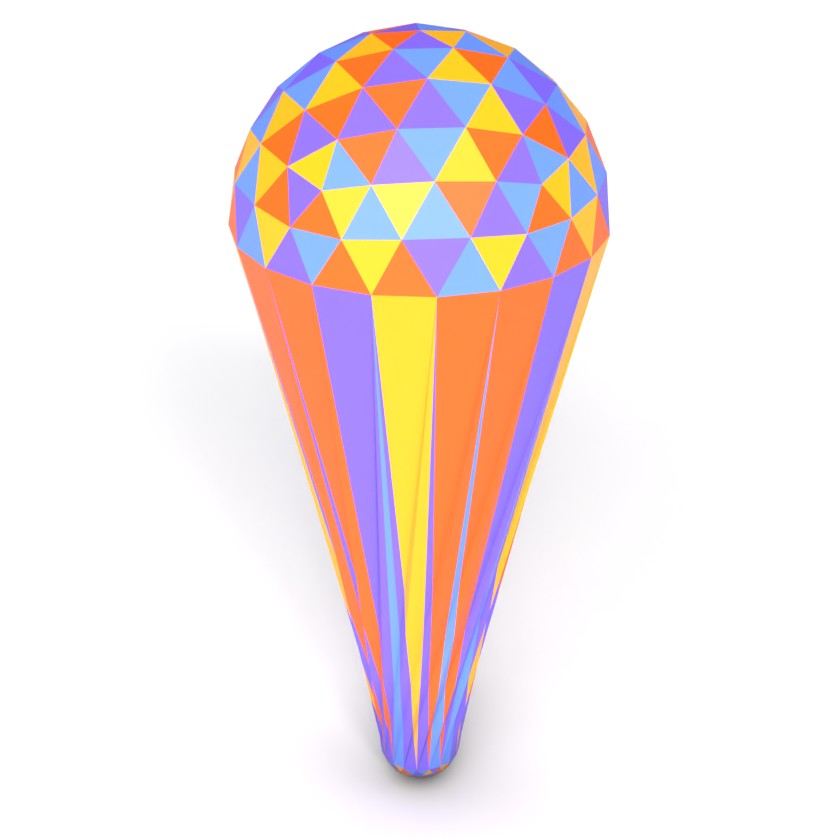}
    \caption{Delaunay}
  \end{subfigure}%
  \hfill%
  \begin{subfigure}{.2\linewidth}
    \centering
    \includegraphics[width=\linewidth]{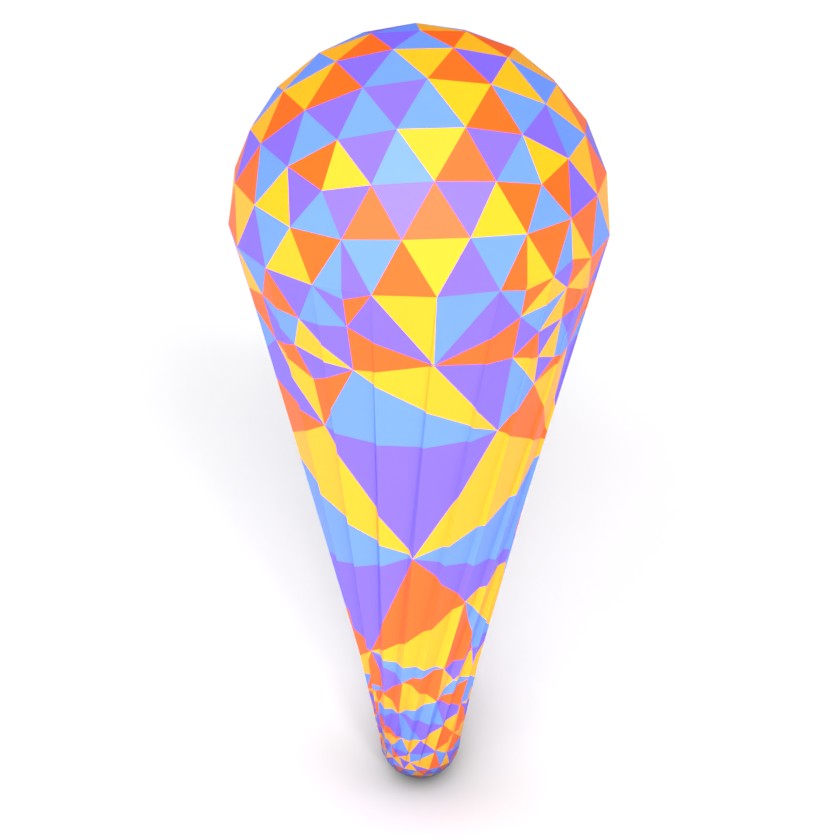}
    \caption{Refined}
  \end{subfigure}%
  \hfill%
  \begin{subfigure}{.2\linewidth}
    \centering
    \includegraphics[width=\linewidth]{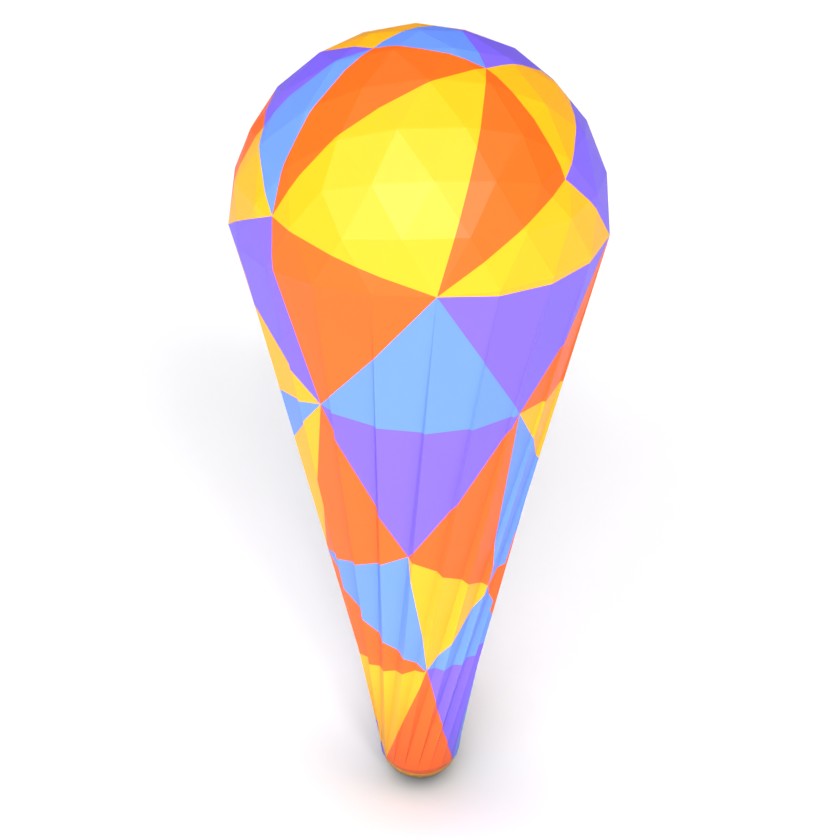}
    \caption{Coarsened}
  \end{subfigure}
  \hfill%
  \hfill%
  \caption{Example illustrating the meshing capabilities of the proposed method. Starting from the initial triangulation (a), the intrinsic approach allows the construction of a Delaunay triangulation (b) as well as refined (c) and coarsened (d) triangulations. The triangulations (b–d) are isogeometric to the initial mesh (a), as they result from its segmentation.}
  \label{fig:cone}
\end{figure*}

To improve surface remeshing methods, geodesic distances computed directly on the surface can be incorporated. This allows meshing operations to account for the actual geometric distances between points. Several methods \cite{peyre2005geodesic, liu2010construction, eck1995multiresolution, floater2002parameterization, peyre2004surface, liu2017constructing} use intrinsic distances to perform more accurate operations such as vertex insertion/repositioning or surface parameterization. However, in most of these approaches, the mesh is still represented with straight-sided edges, which limits the advantages of using geodesic information. In contrast, other methods have been proposed to exploit geodesic distances more extensively. In these approaches, the connection between two points is no longer a straight line in 3D space but a piecewise linear curve computed directly on the surface. As a result, the surface is no longer approximated but is exactly represented by the generated elements.

The first intrinsic methods can be viewed as direct extensions of Delaunay triangulations to piecewise linear surfaces, defining intrinsic Delaunay triangulations \cite{bobenko2007discrete, fisher2007algorithm} and refinement schemes \cite{sharp2019navigating}. In these cases, intrinsic triangulations and all their properties are straightforward to compute because the intrinsic triangles can always be flattened onto the plane. As a result, angles and edge lengths can be computed using standard planar formulas. Moreover, the use of such developable triangles in computation is simple, since they can be treated as standard straight-sided elements without any additional effort. However, the developable nature of these elements imposes significant limitations, as intrinsic triangles alone cannot represent regions with nonzero Gaussian curvature. A direct consequence is that these methods do not allow vertex removal in the mesh and, therefore, can only be used for refinement, not for coarsening. This is particularly unfortunate, as one of the primary goals of meshing is often to reduce the number of elements compared to the input discrete surface. It should be noted that intrinsic coarsening was addressed in \cite{liu2023surface}. However, rather than explicitly computing geodesics, the authors locally modify the geometry by flattening vertices, thereby reducing the problem to a planar configuration in regions where vertex removal is required. This approach results in a mesh in which edges are no longer geodesic.

By adopting a more general framework for intrinsic triangles, it becomes possible to work with non-developable intrinsic triangles. For instance, \cite{xin2011isotropic} constructs a geodesic Delaunay triangulation on sampled surface points to achieve isotropic mesh simplification. Nevertheless, it should be noted that intrinsic Delaunay triangulations do not necessarily exist for an arbitrary set of sampled points on a surface \cite{leibon2000delaunay, liu2017constructing}. In this work, we present an approach for constructing an intrinsic triangulation of a surface that enforces constraints on both element size and quality. This defines a complete intrinsic meshing framework capable of performing both refinement and coarsening. An illustration of these capabilities is provided in Fig.~\ref{fig:cone}, where the different elements are distinguished by various colors. This mesher is based on local optimizations leading to a parametrization-independent and robust method. We also demonstrate how the resulting piecewise-linear elements can be used not only for straight-sided mesh simplification but also for high-order mesh generation.

\subsection{Geodesic Distances}

A central requirement of the proposed intrinsic meshing approach is the
ability to compute geometric quantities directly on the surface. Local
operations require geodesic distances between pairs of points and
intrinsic angles associated with triples of points. These angles 
correspond to the angle on the surface between two geodesics meeting at a point
For Delaunay-type refinement, another key operation is the computation
of geodesic circumcenters. Given three vertices of an intrinsic
triangle, a geodesic circumcenter is a point on the surface that is
equidistant, in geodesic distance, from the three vertices. It therefore
plays the role of the classical Euclidean circumcenter, but in the
intrinsic geometry of the surface.
For developable
intrinsic triangles \cite{sharp2019navigating}, lengths and
angles can be directly inferred from previously
computed information, which significantly simplifies the computation
of these geometric features. In contrast, since our focus here is on
non-developable intrinsic triangles, the approach requires a
mechanism to evaluate geodesic distances on the surface. Once these
distances are obtained, other geometric constructs, such as geodesic
lines, angles and triangle circumcircles, can be derived.

Numerous approximate methods for computing geodesic distances have
been developed in the computer graphics community, many of which are
based on variations of the Fast Marching Method \cite{sethian1996fast,
sethian1999fast}. For our purposes, particularly when dealing with
complex surface geometries, it is crucial to reduce ambiguities and
prevent geodesics from intersecting or becoming indistinguishable.
Moreover, each geodesic must be unique. To achieve this, we employ an
exact geodesic distance computation rather than an approximate one. A
comprehensive survey of both exact and approximate techniques for
computing geodesic distances and paths on surfaces can be found in
\cite{crane2020survey}.

\subsubsection{Exact geodesic distance computation}
The problem of computing shortest paths from a single source to all
destinations on a polyhedral surface was first studied in
\cite{o1985shortest}, where an algorithm with a prohibitive asymptotic
time complexity of $O(n^5)$ was proposed, with $n$ denoting the number
of edges of the surface. A significant advancement was later proposed
in \cite{mitchell1987discrete} with the
Mitchell–Mount–Papadimitriou (MMP) algorithm, which attains a
worst-case time complexity of
$O(n^2 \log n)$ and can be interpreted as a continuous analogue of
Dijkstra’s shortest path algorithm \cite{dijkstra1959note}. An alternative
approach, the Chen–Han (CH) algorithm, was proposed in
\cite{chen1990shortest}, achieving $O(n^2)$ time complexity. However,
empirical comparisons in \cite{xin2009improving} show that the MMP
algorithm significantly outperforms the CH algorithm in practice, both
in terms of computational time and memory usage. In the same study, an
improved version of the CH algorithm, referred to as Improved Chen-Han
(ICH), is introduced. This algorithm sacrifices the theoretical time
complexity advantage of CH by incorporating a priority queue
propagation strategy similar to that of MMP, resulting in time
complexity of $O(n^2 \log n)$. The use of a priority queue in ICH
allows for effective filtering of unnecessary paths, ultimately
producing an algorithm similar to MMP in both time and space
complexity, although ICH shows slightly improved performance in some
cases \cite{xin2009improving}.
Furthermore, \cite{liu2013exact} shows that all these algorithms can
achieve improvements in both time and space by computing distances
using an edge-based structure rather than the classical half-edge
structure. Additionally, CH and ICH can be optimized to reduce memory
usage by restricting the computation of shortest paths to mesh
vertices only.
However, this optimization is not applicable here, since shortest
paths to points on edges or faces are also required. Consequently,
within the scope of this work, the ICH and MMP algorithms exhibit no
significant differences.

\subsubsection{Optimization based on the application}
Significant research has been conducted to improve the performance of
algorithms for the “single-source, all-destinations” problem,
notably through techniques such as parallelization strategies
\cite{ying2014parallel} and grouped information propagation
\cite{xu2015fast, qin2016fast}. While these approaches can accelerate
computation, more effective optimizations are possible when the
objective is to compute the distance to a specific point on the
surface, rather than to all points, as considered in this work. In
\cite{floater2002parameterization}, this is achieved by restricting
propagation to an elliptical region around the source and destination,
using an approximation of the geodesic distance computed via fast
marching methods. This approach was further refined in
\cite{surazhsky2005fast}, where a Dijkstra search limited to edges was
combined with their own approximate method employing merged windows.
The authors also drew inspiration from the A* search algorithm
\cite{hart1968formal}, even using the term “continuous A* search.”
Specifically, they used Euclidean distances as a lower bound on the
distance from a point to the destination. Importantly, these
approximations are still only used to filter which windows are
propagated.

In this work, the Euclidean distance will be used as a lower bound on
the remaining distance from an intermediate point to the destinations.
However, its primary role will be to determine the order in which
windows are propagated, aligning the approach more closely with the
original A* algorithm. This also leads to a considerably simpler
method. A comparable idea is explored in \cite{xin2010applying}
with the ICH algorithm. This method will also be extended to compute a
point that is equidistant from three given points. This will enable
the computation of circumcenters, as in \cite{liu2010construction},
but without requiring recursive mesh refinement. Since a public
implementation of the MMP algorithm is available\footnote{Danil
Kirsanov (2025). Exact geodesic
for triangular meshes (https://www.mathworks.com/matlabcentral
/fileexchange/18168-exact-geodesic-for-triangular-meshes), MATLAB
Central File Exchange. Retrieved May 16, 2025.} from one of the
authors of \cite{surazhsky2005fast}, and because performance
differences between the MMP and ICH algorithms are minor, the
experimental results will be based on this implementation of MMP.
Adapting the approach to the ICH algorithm would require only slight
modifications.

%% file: intrinsic.tex
\section{Intrinsic Meshing}

During the generation of a mesh, the distribution of vertices is governed by a subtle balance: geometry defines where vertices can be placed, the characteristic length determines their spacing, and quality constraints adjust their relative placement. Consequently, resolving all vertex positions in a single step is fundamentally challenging. As a result, most algorithms adopt an iterative approach to progressively modify the mesh toward the desired refinement and quality. In conventional unstructured meshing methods, the fundamental operations are point insertion, point removal and modification of the connectivity of elements. However, in the context considered here, these operations are constrained by the feasibility and computational cost of evaluating the required geometric elements for each operation. As a result, the local optimization operations considered in this work are limited to edge swapping, edge collapse, edge split, and, finally, triangle split.

\input{intrinsicSwap}

\input{intrinsicSplitCollapse}

\input{intrinsicSplitTriangle}

\input{intrinsicOrdering}

%% file: intrinsicSwap.tex
\subsection{Intrinsic Edge Swap}

\begin{figure}[t]
  \centering
  \begin{tikzpicture}[x=0.1\linewidth, y=0.1\linewidth]
    \node (A) at (0, 0) [inner ysep=0pt] {
      \includegraphics[width=0.175\linewidth]{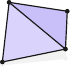}
    };
    \node (B) at (5, 1.2) [inner ysep=0pt] {
      \includegraphics[width=0.175\linewidth]{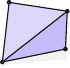}
    };
    \node (C) at (5, -1.2) [inner ysep=0pt] {
      \includegraphics[width=0.175\linewidth]{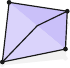}
    };
    \draw[-latex, line width=2] (A) -- (B) node[midway, above, sloped] {Extrinsic};
    \draw[-latex, line width=2] (A) -- (C) node[midway, below, sloped] {Intrinsic};
  \end{tikzpicture}
  \caption{Intrinsic and extrinsic edge swap.}
  \label{fig:swap}
\end{figure}

Given a set of points sampled on a surface, there exist multiple valid triangulations that can be constructed. Here, “valid” refers to triangulations composed of non-overlapping, non-intersecting triangles that preserve the topology of the underlying surface. However, not all triangulations exhibit the same properties in terms of field interpolation and numerical computation. In particular, nearly degenerate triangles tend to extrapolate values far from their vertices, leading to increased approximation errors and reduced numerical stability.

\subsubsection{Delaunay Triangulation}
In dimension $d$, the Delaunay triangulation of a set of points is a
triangulation of their convex hull such that the circumsphere of each
$d$-simplex contains no other point in its interior. In two dimensions,
this condition reduces to the empty circumcircle property for triangles.
In low dimensions, Delaunay triangulations
can be constructed efficiently using well-established algorithms,
which explains their central role in practical meshing methods.
The planar Delaunay triangulation is particularly important for mesh
generation because it satisfies a max-min angle property: among all
triangulations of the same point set, it maximizes the smallest angle.
This tends to avoid skinny triangles, which is beneficial for numerical
interpolation and simulation. 
In two dimensions, another important property is that the Delaunay
triangulation can be reached from any triangulation of the same point set
by a finite sequence of local edge swaps. More precisely, if $n$ denotes
the number of vertices, at most $O(n^2)$ edge swaps are required in the
worst case.

\subsubsection{Intrinsic Delaunay Triangulation}
An intrinsic Delaunay triangulation is a triangulation whose vertices lie
on a surface and whose edges are represented by geodesic paths on that
surface, such that each triangle satisfies an empty geodesic circumcircle
condition. More precisely, for every intrinsic triangle, there exists a
point on the surface that is equidistant, in geodesic distance, from its
three vertices, and the corresponding geodesic disk contains no other
vertex of the triangulation in its interior. The classical two-dimensional 
edge-flipping algorithm suggests a natural
extension to surfaces. Instead of flipping straight edges in the plane,
one can perform intrinsic edge swaps, where the new edge is recomputed
as a geodesic path on the surface. Each swap is driven by a local
intrinsic Delaunay criterion: if an edge is not locally Delaunay, the
algorithm attempts to replace it by the opposite geodesic edge. In this
way, the mesh is iteratively modified to enforce local Delaunayness,
while remaining entirely defined on the surface. Fig.~\ref{fig:swap} illustrates the 
difference between a classical (extrinsic) edge swap and an intrinsic one.

In this work, mesh edges are defined as the shortest geodesic paths between vertices. This choice 
eliminates the ambiguity arising from the existence of multiple geodesics connecting the same pair of 
vertices and provides a robust framework for computing these geodesics consistently. However, restricting 
edges to shortest paths also limits the set of representable geodesics. Consequently, constructing 
intrinsic meshes based solely on shortest paths imposes constraints on the admissible triangulations, 
as discussed in Sec.~\ref{sec:swapfeasibility} and~\ref{sec:insertDelaunay}.

\subsubsection{Edge Swapping Feasibility}\label{sec:swapfeasibility}

\begin{figure}[t]
  \centering
  \begin{subfigure}{.5\linewidth}
    \centering
    \includegraphics[width=.75\linewidth]{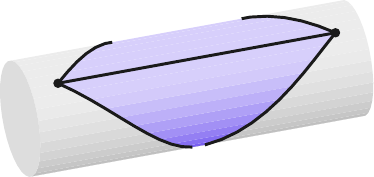}
  \end{subfigure}%
  \begin{subfigure}{.5\linewidth}
    \centering
    \includegraphics[width=.75\linewidth]{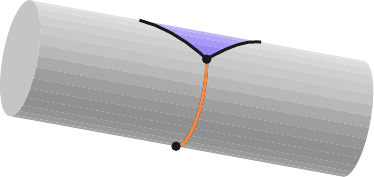}
  \end{subfigure}
  \caption{Example of an edge that cannot be swapped because the candidate edge does not lie within the cavity of the adjacent triangles.}
  \label{fig:cannotswap}
\end{figure}

In the planar setting, it is always possible to swap an edge that is not locally Delaunay, thus verifying that an edge should be swapped is sufficient to guarantee that the operation is feasible. On curved surfaces, this property no longer holds. Indeed, the resulting edge after a swap may fail to remain within the cavity formed by the two adjacent triangles, as it is the case in Fig.~\ref{fig:cannotswap}. It is therefore essential to verify that the candidate edge lies entirely within this cavity. Moreover, to preserve the validity of the mesh, edges must remain non-intersecting except at shared vertices. As described in Alg.~\ref{alg:canWeSwap}, both conditions can be enforced through an edge-intersection query: first, by confirming that the current and candidate edges do intersect, and second, by verifying that the candidate edge does not intersect the boundary edges of the cavity.

\begin{algorithm}
  \caption{\textsc{canWeSwap}($e$)}
  \label{alg:canWeSwap}
  \begin{algorithmic}[1]
    \STATE Let $o$ be the edge opposite to $e$
    \IF{$e$ and $o$ do not intersect}
      \RETURN \FALSE
    \ENDIF
    \STATE Construct the set of edges surrounding $e$ and $o$: $E_{\text{border}}$
    \IF{\textsc{checkIntersections}(\{$o$\}, $E_{\text{border}}$)}
      \RETURN \FALSE
    \ENDIF
    \RETURN \TRUE
  \end{algorithmic}
\end{algorithm}

\subsubsection{Edge Swapping Heuristic Criterion}

\begin{figure}[b]
  \centering
  \begin{subfigure}{.5\linewidth}
    \centering
    \includegraphics[width=.75\linewidth]{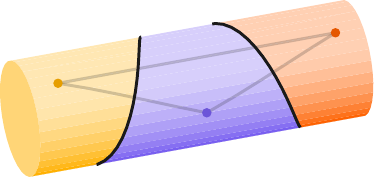}
  \end{subfigure}%
  \begin{subfigure}{.5\linewidth}
    \centering
    \includegraphics[width=.75\linewidth]{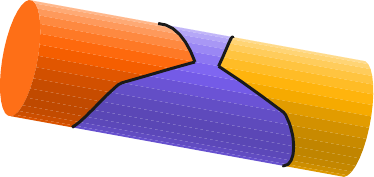}
  \end{subfigure}
  \caption{Example of an intrinsic triangle with no circumcircle.}
  \label{fig:uniqueandexist}
\end{figure}

The Delaunay criterion can also be directly generalized to non-planar surfaces by replacing circumcircles with geodesic distance-based circumcircles. For each edge, one could compute the circumcenters of both adjacent triangles and check whether the Delaunay criterion is locally satisfied. However, the geodesic circumcenter of a triangle on a curved surface is neither guaranteed to be unique nor to even exist, see Fig.~\ref{fig:uniqueandexist}. Moreover, even if it exists, its computation might be heavy and also might not be possible due to numerical precision limitations as it will be discussed in Sec.~\ref{sec:circumcenter}. As a consequence, the geodesic Delaunay criterion is not employed as it is. Instead, a heuristic based on readily available information will be adopted.

The proposed heuristic is required to satisfy several key properties. First, it must be consistent with the Delaunay criterion for developable triangles, which corresponds to the planar scenario. Second, it must ensure coherence, preventing newly introduced edges from being immediately swapped back. Third, it should not presuppose that the candidate edge resulting from a swap lies within the cavity, as this condition does not always hold. Finally, the heuristic must depend exclusively on intrinsic quantities simple to compute, specifically the edge lengths and angles defined on the surface.

In the planar case, a simple implementation of the Delaunay criterion consists in evaluating the two angles opposite to the edge under consideration and checking whether their sum exceeds $\pi$. If this condition is satisfied, the edge is flipped. Equivalently, this procedure can be interpreted as minimizing the sum of the angles opposite to edges, since the total sum of angles in a quadrilateral is equal to $2\pi$. Motivated by this observation, the heuristic proposed in this work selects the configuration that minimizes the sum of the angles opposite to the edge. This approach does not rely on any assumptions regarding the candidate edges and depends only on intrinsic angles.

\begin{algorithm}
\caption{\textsc{doWeSwap}($e$)}
  \label{alg:doWeSwap}
\begin{algorithmic}[1]
\STATE Compute opposite angles $\alpha, \alpha'$ of edge $e$
\STATE Compute opposite angles $\beta, \beta'$ of the opposite edge of $e$
\RETURN $\alpha + \alpha' > \beta + \beta'$
\end{algorithmic}
\end{algorithm}

\begin{figure}[t]
  \centering
  \hfill%
  \begin{subfigure}{.4\linewidth}
    \centering
    \includegraphics[width=.9\linewidth]{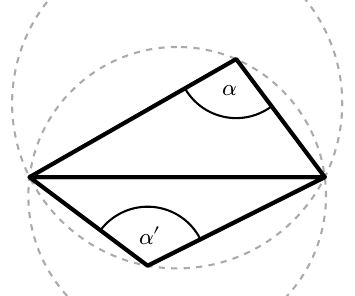}
    \caption{Not locally Delaunay}
  \end{subfigure}%
  \hfill%
  \begin{subfigure}{.4\linewidth}
    \centering
    \includegraphics[width=.9\linewidth]{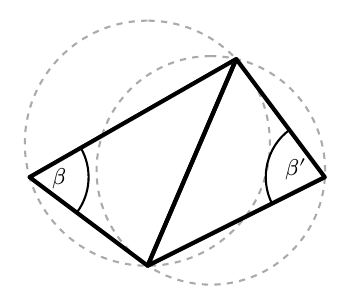}
    \caption{Locally Delaunay}
  \end{subfigure}
  \hfill%
  \hfill%
  \caption{Delaunay criterion in the plane given as the inequality on opposite angles: $\alpha + \alpha' > \pi > \beta + \beta'$.}
\end{figure}

\subsubsection{Unreachable Delaunay Configurations}\label{sec:insertDelaunay}

\begin{figure}[t]
  \centering
  \hfill%
  \begin{subfigure}{.4\linewidth}
    \centering
    \includegraphics[width=\linewidth]{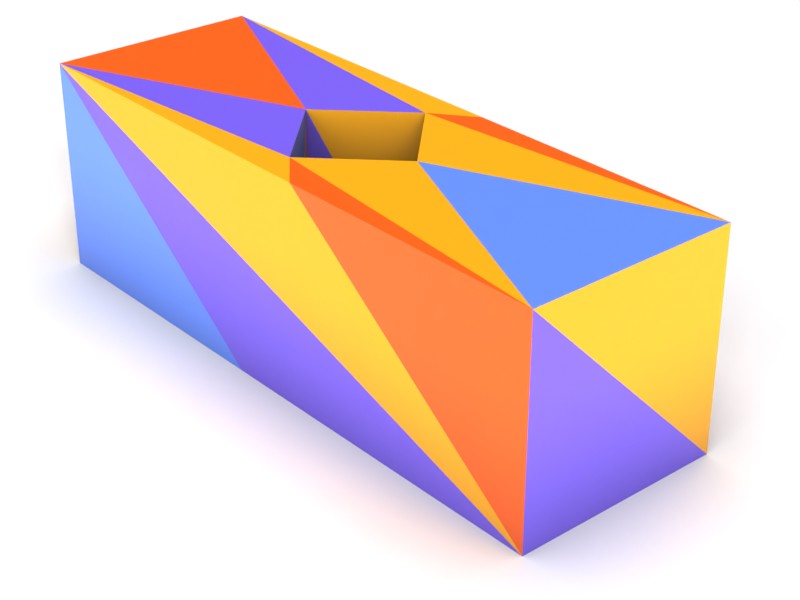}
    \caption{Without point insertion}
    \label{fig:noinsertdelaunay}
  \end{subfigure}%
  \hfill%
  \begin{subfigure}{.4\linewidth}
    \centering
    \includegraphics[width=\linewidth]{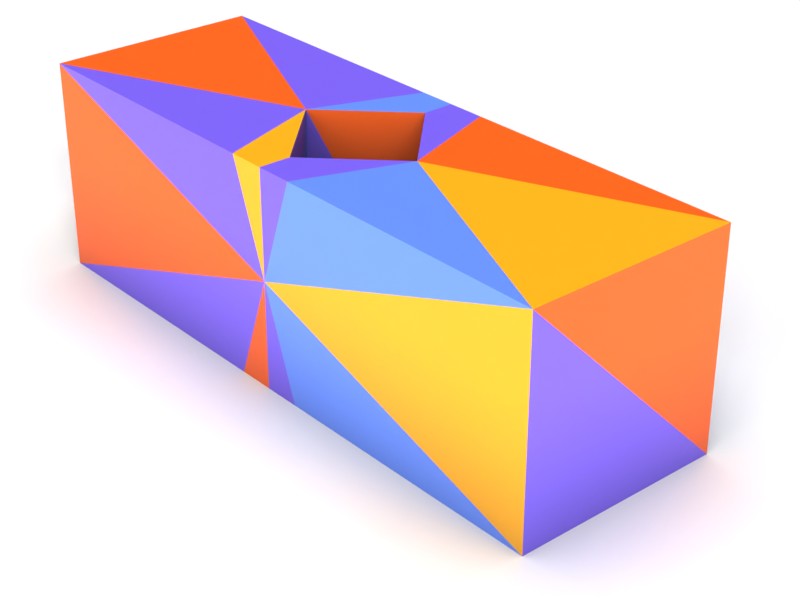}
    \caption{With point insertion}
    \label{fig:insertDelaunay}
  \end{subfigure}
  \hfill%
  \hfill%
  \caption{Intrinsic edge swaps of a box with a hole without and with point insertion. A Delaunay configuration could not be reached without point insertion.}
\end{figure}

As it was discussed in Sec.~\ref{sec:swapfeasibility}, some configurations prohibit an edge flip when the resulting geodesic would lie outside the cavity or intersect other edges, thereby altering the mesh validity. In such cases, reaching a Delaunay configuration may become impossible if the swap cannot be avoided. An illustration of this situation is provided in Fig.~\ref{fig:noinsertdelaunay}.

A possible strategy to recover a Delaunay configuration consists in splitting edges that would otherwise be swapped but cannot be flipped due to topological constraints. Although this operation introduces an additional vertex, it locally modifies the mesh connectivity, thereby enabling the construction of an intrinsic mesh in situations where the surface topology would otherwise prevent it. As illustrated in Fig.~\ref{fig:insertDelaunay}, inserting a vertex at the midpoint of such problematic edges is sufficient to resolve the issue and reach a Delaunay configuration.

\subsubsection{Existence and Uniqueness of Intrinsic Delaunay Triangulation}\label{sec:existandunique}

Beyond the limitations in representing certain Delaunay configurations, there exist situations in which a valid Delaunay configuration may fail to exist due to the topology of the underlying surface. This issue is independent of the representation of intrinsic triangles and is inherently topological, as discussed in \cite{liu2014duality}. In that work, the existence of a valid Delaunay triangulation is established under a condition on the dual Voronoi diagram, namely the closed ball property: (1) each Voronoi cell is homeomorphic to a planar disk, (2) the intersection of any two Voronoi cells contains at most one Voronoi edge, and (3) the intersection of any three Voronoi cells contains at most one Voronoi vertex.

Moreover, a constructive procedure is proposed to enforce the existence of a Delaunay triangulation by iteratively inserting additional points along Voronoi edges and pseudo-bisectors, i.e., loci of points admitting multiple shortest paths to a Voronoi vertex. Although the point insertion strategy described in Sec.~\ref{sec:insertDelaunay} differs from that of \cite{liu2014duality}, it is reasonable to expect that the existence of a Delaunay triangulation can still be enforced, provided that a sufficient number of points are inserted.

\subsubsection{Edge Swapping Algorithm}

In general, the order in which edges are swapped does not affect the final result, as the sequence of swaps converges to a unique Delaunay configuration. However, in the present context, additional points may be inserted to enforce the existence of such a configuration. Consequently, the order in which points are inserted can influence the final triangulation, potentially leading to different configurations.

Since there is no natural or canonical strategy to order the edge swaps and point insertions, we adopt a simple deterministic approach. Specifically, edge swaps are performed according to a predefined ordering (e.g., based on vertex numbering), ensuring reproducible behavior of the algorithm.

The complete edge-swapping procedure is detailed in Alg.~\ref{alg:swapEdges}. The function \textsc{doWeSwap} applies the heuristic Delaunay criterion to the neighboring triangles, while \textsc{canWeSwap} checks for intersections involving the candidate edge, as described in Alg.~\ref{alg:canWeSwap} and \ref{alg:doWeSwap}.

\begin{algorithm}
  \caption{\textsc{swapEdges}($allowInsert$)}
  \label{alg:swapEdges}
  \begin{algorithmic}[1]
    \STATE Initialize a set $S$ of all edges
    \WHILE{$S \neq \emptyset$}
      \STATE Extract an edge $e$ from $S$
      \IF{\NOT \textsc{doWeSwap}($e$)}
        \STATE \textbf{continue}
    \ENDIF
    \IF{\textsc{canWeSwap}($e$)}
      \STATE Swap edge $e$
        \STATE $S \leftarrow S \cup \{ \text{adjacent edges} \}$
        \STATE \textbf{continue}
    \ENDIF
    \IF{\NOT $allowInsert$}
        \STATE \textbf{continue}
    \ENDIF   
    \IF {\textsc{trySplitEdge}($e$)}
        \STATE $S \leftarrow S \cup \{ \text{adjacent edges} \}$
    \ENDIF
    \ENDWHILE
  \end{algorithmic}
\end{algorithm}

%% file: intrinsicSplitCollapse.tex
\subsection{Intrinsic Edge Split and Collapse}

\begin{figure}[b]
  \centering
  \begin{tikzpicture}[x=0.1\linewidth, y=0.1\linewidth]
    \node (A) at (0, 0) {
      \includegraphics[width=0.2\linewidth]{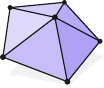}
    };
    \node (B) at (5, +1.2) {
      \includegraphics[width=0.2\linewidth]{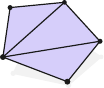}
    };
    \node (C) at (5, -1.2) {
      \includegraphics[width=0.2\linewidth]{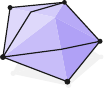}
    };
    \draw[-latex, line width=2] (A) -- (B) node[midway, above, sloped] {Extrinsic};
    \draw[-latex, line width=2] (A) -- (C) node[midway, below, sloped] {Intrinsic};
  \end{tikzpicture}
  \caption{Intrinsic and extrinsic edge collapse.}
  \label{fig:collapse}
\end{figure}

A fundamental requirement for meshing algorithms is the capability to enforce spatially varying refinement across the geometry, commonly specified via a characteristic length field. The objective of the meshing process is then to produce edges whose dimensionless lengths are close to unity, so that regions requiring finer resolution correspond to smaller prescribed characteristic lengths. This is accomplished by locally adjusting element sizes.

Edge refinement can be achieved by splitting an edge, which replaces two triangles with four, thereby reducing local edge lengths. Conversely, edge coarsening can be performed by collapsing an edge into a single vertex, eliminating two triangles and updating the connectivity of adjacent elements. These operations are inherently local, as they only require the computation of the new vertex and its incident edges, and they naturally generalize to the intrinsic setting, see Fig.~\ref{fig:collapse}. Following each split or collapse, the restoration of a Delaunay configuration involves testing edge flips only within the modified neighborhood. These local edge operations are therefore employed to enforce the prescribed element size constraints.

\subsubsection{Edge Split and Collapse Criterion}

The decision to split or collapse an edge is based solely on its dimensionless length relative to a prescribed target length. Since this target value cannot, in general, be satisfied exactly, it is necessary to define an admissible range within which element sizes are considered acceptable. In this context, the minimum and maximum allowable characteristic lengths must be chosen to be at most $1/\sqrt{2}$ and at least $\sqrt{2}$ times the target length, respectively, in order to prevent immediate reversal of operations (i.e., collapsing an edge directly after splitting it, and vice versa). This condition, however, remains weak: even when it is satisfied, certain configurations may still lead to oscillatory sequences of splits and collapses. Consequently, in practice, wider tolerances are typically required to ensure robust behavior.

To provide greater flexibility, the characteristic length in this work is defined as a range of admissible element sizes, specified by a minimum and a maximum characteristic length. A single target value (e.g., $l_c$) may still be prescribed and systematically converted into an interval by introducing fixed margins (e.g., $[0.5 l_c, 2 l_c]$). Additionally, the formulation allows for relaxed constraints by omitting bounds on the minimum and/or maximum element size. The primary limitation of this approach is that, if the admissible range is too narrow, the method may be unable to satisfy all constraints simultaneously and consequently fail to converge.

\begin{algorithm}
\caption{\textsc{doWeSplitEdge}($e$)}
  \label{alg:doWeSplitEdge}
\begin{algorithmic}[1]
    \STATE Let $cl_{\text{max}}(\mathbf{x})$ be the maximum characteristic length
  \STATE $l_{\text{max}} \gets \int_e dl / cl_{\text{max}}$
\RETURN $l_{\text{max}} > 1$
\end{algorithmic}
\end{algorithm}

\begin{algorithm}
\caption{\textsc{doWeCollapseEdge}($e$)}
  \label{alg:doWeCollapseEdge}
\begin{algorithmic}[1]
    \STATE Let $cl_{\text{min}}(\mathbf{x})$ be the minimum characteristic length
  \STATE $l_{\text{min}} \gets \int_e dl / cl_{\text{min}}$
\RETURN $l_{\text{min}} < 1$
\end{algorithmic}
\end{algorithm}

\subsubsection{Edge Split Feasibility}

When splitting an edge, a first verification consists in ensuring that the two newly created edges lie entirely within the cavity, i.e., that they do not intersect any boundary edges. In addition, it must be verified that these new edges do not intersect each other.

After the split operation, the mesh must be restored to a Delaunay configuration. To this end, adjacent edges are iteratively examined and flipped when necessary, following a procedure similar to Alg.~\ref{alg:swapEdges}, but restricted to the local neighborhood. If, during this process, an edge cannot be flipped and cannot be further subdivided, the operation is aborted. This limitation implies that, in certain configurations, it may not be possible to enforce the prescribed size constraints through edge splitting alone.

\subsubsection{Edge Collapse Feasibility}

The constraints on edge collapses are similar to those for edge splits but are generally stricter. Since the purpose of an edge collapse is to increase the local mesh size by removing a vertex, the operation is aborted if the subsequent edge-flipping phase introduces a new vertex, as this would contradict the intended coarsening.

Additionally, a topological check must be performed to ensure the mesh remains valid after the collapse. Specifically, if the cavity in which the edge is to be collapsed is not a topological disk, the collapse would alter the mesh topology and render it invalid as discussed in Sec.~\ref{sec:existandunique}. This condition can be efficiently verified using the triangles and boundary edges of the cavity, as described in Alg.~\ref{alg:inDiskCavity}.

\begin{algorithm}
  \caption{\textsc{inDiskCavity}($e$)}
  \label{alg:inDiskCavity}
  \begin{algorithmic}[1]
    \STATE $\mathcal{T} \gets \text{triangles in cavity of } e$
    \STATE $\mathcal{B} \gets \text{border edges of cavity of } e$
    \IF{$|\mathcal{T}| \neq |\mathcal{B}| + 2$}
      \RETURN \FALSE
    \ENDIF
    \FORALL{distinct pairs of border edges $b_0, b_1$ of $\mathcal{B}$}
      \IF{$b_0$ and $b_1$ have the same first vertex \OR $b_0$ and $b_1$ have the same second vertex}
        \RETURN \FALSE
      \ENDIF
    \ENDFOR
    \RETURN \TRUE
  \end{algorithmic}
\end{algorithm}

When collapsing an edge, the new vertex can be placed anywhere within the cavity. However, selecting one of the edge’s endpoints is preferable, as it enables reuse of previously computed geodesics. Also, permitting the edge to collapse to its midpoint can facilitate simplification in cases where collapsing to either of its endpoints is not feasible. Subsequently, the selected point is the one that maximizes a quality criterion that will be later defined in Sec.~\ref{sec:quality}. After the edge collapse, adjacent edges must also be checked for possible flips in order to restore a Delaunay configuration. This operation may affect a region larger than the local cavity, since the midpoint can be inserted and the Delaunay criterion is enforced only heuristically.

\subsubsection{Unstable Delaunay Configurations}

\begin{figure}[htbp]
  \centering
  \begin{tikzpicture}[>=latex]
  \def\r{2.2}
  \foreach \i in {0,...,4} {
      \node (img\i) at ({-72*(\i)}:\r)
      {\includegraphics[width=2cm, trim={10cm 12cm 10cm 3cm}, clip]{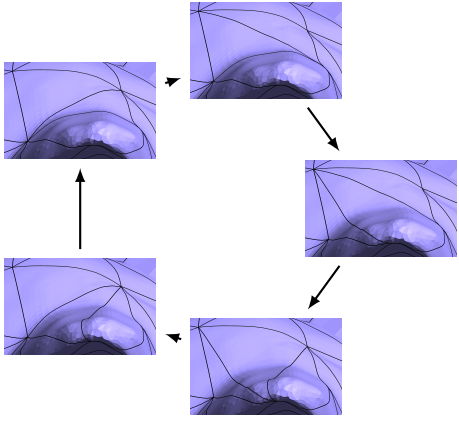}};
  }
  \foreach \i in {0,...,4} {
      \pgfmathtruncatemacro{\next}{mod(\i+1,5)}
      \draw[->, thick] (img\i) -- (img\next);
  }
  \end{tikzpicture}
  \caption{Configurations where the heuristic swap condition does not lead to a stable configuration.}
  \label{fig:endless}
\end{figure}

When modifying an intrinsic mesh through vertex insertion or removal, the edge-flipping algorithm may fail to converge to a stable configuration. This limitation arises from the heuristic nature of the Delaunay criterion, which can permit repeated edge swaps and lead to infinite cycling, as illustrated in Fig.~\ref{fig:endless}. In practice, such situations can be detected by imposing a limit on the number of swaps, allowing the operation to be safely rejected.

\subsubsection{Edge Split and Collapse Algorithms}

Unlike edge swapping, the order of edge splits and collapses is crucial. Arbitrary splitting can lead to oversampling and may affect convergence, whereas uncontrolled collapsing can prevent further operations, limiting the method’s coarsening capabilities. To address this, edge splitting should begin with the longest edges, while edge collapsing should start with the shortest.

These operations are implemented using a queue of edges sorted by length. After a split or collapse, both the affected edge and any edges no longer in the mesh must be removed from this queue, and any newly created edges eligible for further operations must be inserted. The full edge splitting and collapsing procedures are detailed in Alg.~\ref{alg:splitEdges} and \ref{alg:collapseEdges}.

\begin{algorithm}
  \caption{\textsc{splitEdges}()}
  \label{alg:splitEdges}
  \begin{algorithmic}[1]
    \STATE Initialize an ordered set $Q \gets \emptyset$
    \FORALL{edges $e$}
      \IF{\textsc{doWeSplitEdge}($e$)}
        \STATE $Q \gets Q \cup e$
      \ENDIF
    \ENDFOR
    \WHILE{$Q \neq \emptyset$}
      \STATE Extract the longest edge $e$ from $Q$
      \IF{\textsc{trySplitEdge($e$)}}
        \FORALL{removed edges $r$}
          \STATE $Q \gets Q \setminus r$
        \ENDFOR
        \FORALL{adjacent edges $a$}
          \IF{\textsc{doWeSplitEdge}($a$)}
              \STATE $Q \gets Q \cup a$
          \ENDIF
        \ENDFOR
      \ENDIF
    \ENDWHILE
  \end{algorithmic}
\end{algorithm}

\begin{algorithm}
  \caption{\textsc{collapseEdges}()}
  \label{alg:collapseEdges}
  \begin{algorithmic}[1]
    \STATE Initialize an ordered set $Q \gets \emptyset$
    \FORALL{edges $e$}
      \IF{\textsc{doWeCollapseEdge}($e$)}
        \STATE $Q \gets Q \cup e$
      \ENDIF
    \ENDFOR
    \WHILE{$Q \neq \emptyset$}
      \STATE Extract the shortest edge $e$ from $Q$
      \IF{\textsc{tryCollapseEdge($e$)}}
        \FORALL{removed edges $r$}
          \STATE $Q \gets Q \setminus r$
        \ENDFOR
        \FORALL{adjacent edges $a$}
          \IF{\textsc{doWeCollapseEdge}($a$)}
              \STATE $Q \gets Q \cup a$
          \ENDIF
        \ENDFOR
      \ENDIF
    \ENDWHILE
  \end{algorithmic}
\end{algorithm}

%% file: intrinsicSplitTriangle.tex
\subsection{Intrinsic Triangle Split}

\begin{figure}[htbp]
  \centering
  \begin{subfigure}{.5\linewidth}
      \centering
      \includegraphics[width=\linewidth]{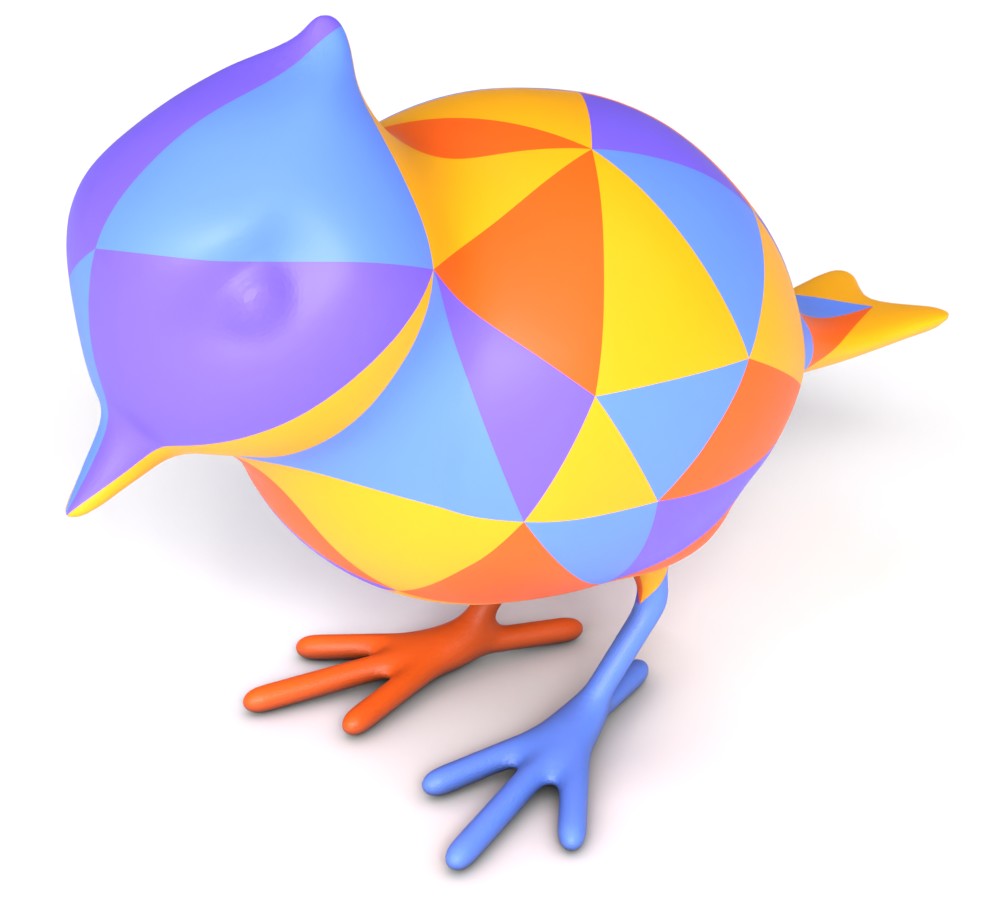}

  \end{subfigure}%

  \hfill%
  \begin{subfigure}{.2\linewidth}
      \centering
      \includegraphics[width=\linewidth]{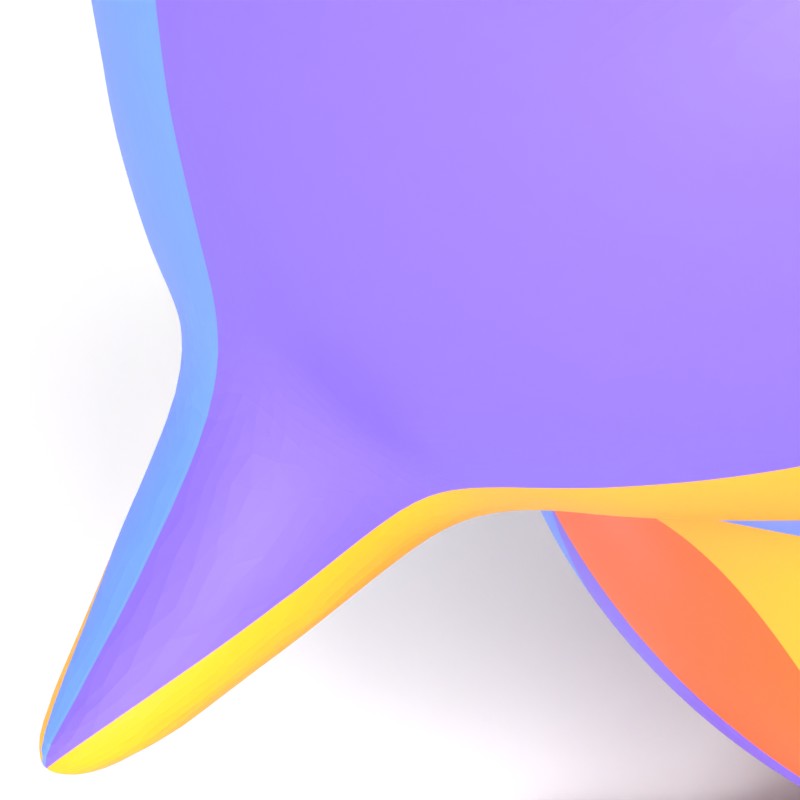}
  \end{subfigure}%
  \hfill%
  \begin{subfigure}{.2\linewidth}
      \centering
      \includegraphics[width=\linewidth]{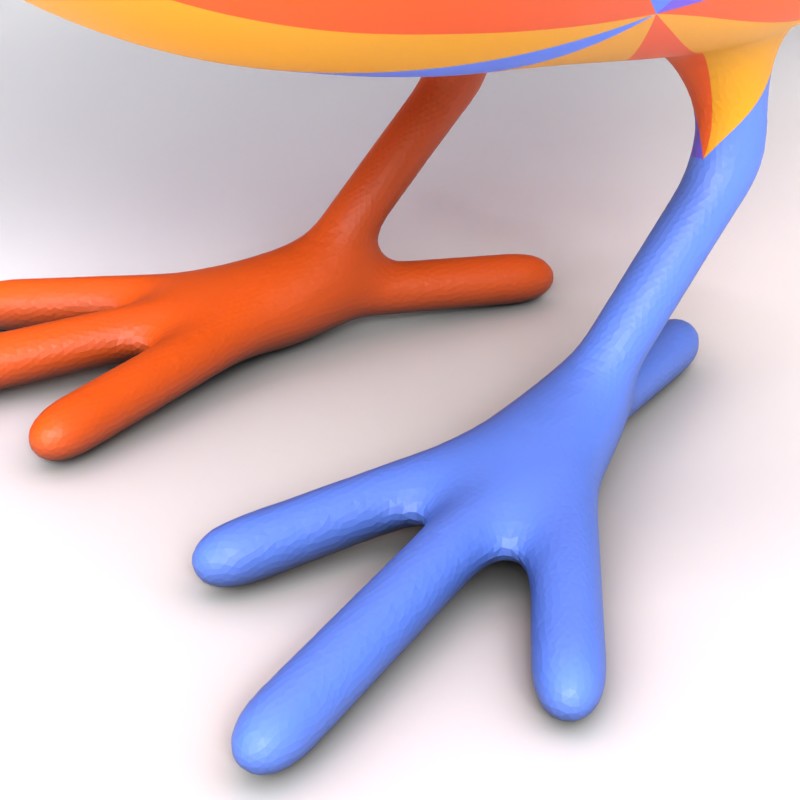}
  \end{subfigure}%
  \hfill%
  \begin{subfigure}{.2\linewidth}
      \centering
      \includegraphics[width=\linewidth]{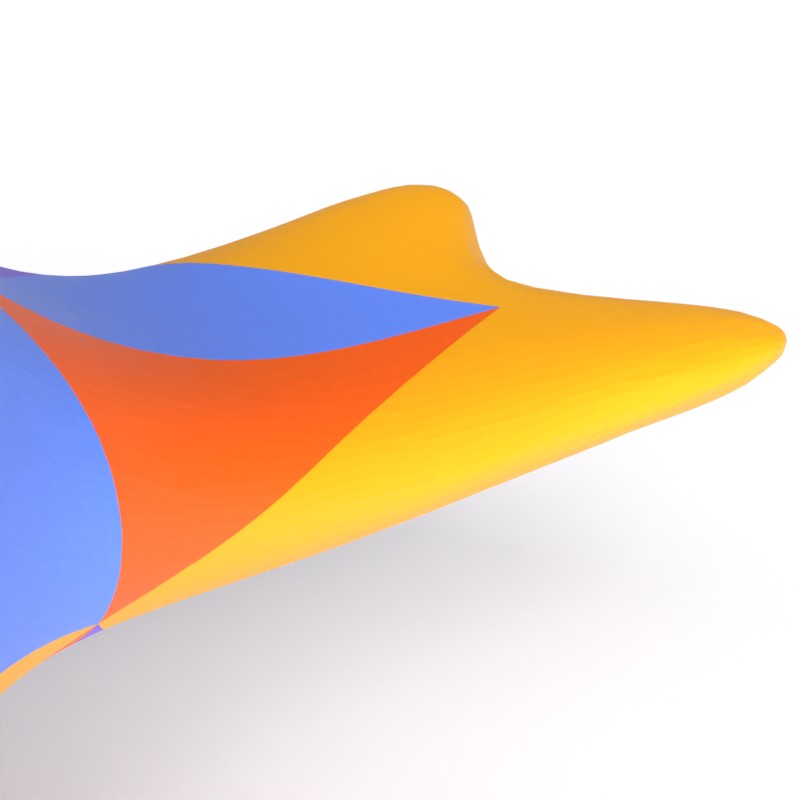}
  \end{subfigure}%
  \hfill%
  \hfill%

  \caption{Example of intrinsic mesh without the split triangle operation (Thingi10k model ID: 178340).}
  \label{fig:swapSplitCollapse}
\end{figure}

Using only edge swap, split, and collapse operations, a mesher can achieve its primary goals. Splits and collapses adjust edges toward the target size defined by characteristic lengths, while swaps improve triangulation quality via a Delaunay-like criterion. An example intrinsic mesh is shown in Fig.~\ref{fig:swapSplitCollapse}. For some applications, this mesh may be sufficient, but additional control may be required to constrain other mesh properties.

Some triangles remain challenging to approximate in regions with concentrated geometric details. Even when edges meet the target lengths, triangle areas can remain large, as sizing depends solely on edge lengths. A quality criterion is therefore needed to identify such triangles, and a triangle split operation is introduced to refine them.

\subsubsection{Triangle Quality}\label{sec:quality}

In conventional meshes, triangle angles are the primary measure of element quality, as extreme angles can significantly degrade computational performance. Other metrics, such as aspect ratio or circumradius-to-shortest-edge ratio, also provide valuable information but are ultimately determined by the angles in the plane. In intrinsic meshes, however, these relationships no longer hold: different quality measures can convey complementary information.

Nonetheless, angle-based quality remains a useful and straightforward indicator. Metrics involving the circumcircle or the area of an intrinsic triangle are generally impractical, as their computation is costly. For simplicity, we therefore define a quality $q_0$ based on the intrinsic angle $\theta$, analogous to the planar case:
\begin{equation}
  q_0(\theta) =  \frac{3}{\pi} \min\left(\theta, \frac{\pi - \theta}{2}\right)
\end{equation}
where $q_0 = 1$ for $\theta=\pi/3$ and $q_0$ becomes negative for angles that are negative or greater than $\pi$. The quality of a triangle is then defined as the minimum of the quality measures associated with its interior angles.

\subsubsection{Triangle Split Criterion}

In addition to maximizing triangle quality, it is desirable to enforce a lower bound ensuring that all elements satisfy a minimum quality requirement. To this end, given prescribed minimum and maximum angles $\theta_{\text{min}}$ and $\theta_{\text{max}}$, we define a quality measure $q$ such that negative $q$ are considered invalid and therefore the corresponding triangle must be refined:
\begin{equation}
q(\theta) = \min\left(
    \frac{\theta - \theta_{\text{min}}}{\pi/3 - \theta_{\text{min}}},
    \frac{\theta_{\text{max}} - \theta}{\theta_{\text{max}} - \pi/3}
\right)
\end{equation}

When enforcing a mesh quality criterion, all operations that could violate it must be prevented. For a triangle-based criterion, collapse and split operations must include additional checks to avoid creating low-quality triangles, as detailed in Alg.~\ref{alg:trySplitEdge} and \ref{alg:tryCollapseEdge}. Similarly, for an edge-based criterion, edge swaps would also need to be monitored to prevent the introduction of poorly shaped edges.

\begin{algorithm}
  \caption{\textsc{quality}($t$)}
  \label{alg:quality}
  \begin{algorithmic}[1]
    \STATE $q^{t} \gets 1$
    \FORALL{angles of $t$}
      \STATE Compute associated quality $q$
      \STATE $q^t \gets \min(q^t, q)$
    \ENDFOR
    \RETURN $q^t$
  \end{algorithmic}
\end{algorithm}

\begin{algorithm}
  \caption{\textsc{doWeSplitTriangle}($t$)}
  \label{alg:doWeSplitTriangle}
  \begin{algorithmic}[1]
    \RETURN \textsc{quality}($t$) $< 0$
  \end{algorithmic}
\end{algorithm}

\subsubsection{Triangle Split Feasibility}

When splitting a triangle, a new vertex is introduced to eliminate it. The placement of this vertex is critical: the removed triangle must not reappear after subsequent edge swaps, and the insertion should avoid the creation of poorly shaped elements. Classical Delaunay refinement methods \cite{ruppert1995delaunay, chew1993guaranteed} place the vertex at the circumcenter of the triangle. This choice guarantees, by the Delaunay criterion, that the triangle will not reappear, while also maximizing the distance to existing vertices, thereby promoting good element quality. In this context, the computational cost associated with evaluating the intrinsic circumcircle is considered acceptable.

A practical difficulty arising from this choice is that the circumcenter may lie outside the triangle to be removed, requiring the identification of the triangle or edge that contains it. Additional edge swaps are then necessary to recover a Delaunay configuration in which the target triangle is effectively removed. Consequently, intersection checks are performed both after the insertion and during the sequence of edge swaps, similarly to the procedures used for edge split and collapse operations.

Although a Delaunay triangulation is guaranteed to exist, practical limitations may arise. Numerical issues can prevent the computation of circumcenters, and the heuristic Delaunay criterion may fail to eliminate the targeted triangle after edge swaps. Consequently, triangle splitting is not always successful.

\subsubsection{Triangle Split Algorithm}

In conventional Delaunay refinement, splitting triangles in order of increasing quality, starting with the worst, has been shown to reduce the total number of mesh elements \cite{shewchuk2002delaunay}. A similar approach is adopted here, using an ordered triangle-splitting procedure as outlined in Alg.~\ref{alg:splitTriangles}.

\begin{algorithm}
  \caption{\textsc{splitTriangles()}}
  \label{alg:splitTriangles}
  \begin{algorithmic}[1]
    \STATE Initialize an ordered set $Q \gets \emptyset$
    \FORALL{triangles $t$}
      \IF{\textsc{doWeSplitTriangle}($t$)}
        \STATE $Q \gets Q \cup t$
      \ENDIF
    \ENDFOR
    \WHILE{$Q \neq \emptyset$}
      \STATE Extract the triangle $t$ with the worst quality from $Q$
      \IF{\textsc{trySplitTriangle($t$)}}
        \FORALL{removed triangles $r$}
          \STATE $Q \gets Q \setminus r$
        \ENDFOR
        \FORALL{created triangles $c$}
          \IF{\textsc{doWeSplitTriangle}($c$)}
              \STATE $Q \gets Q \cup c$
          \ENDIF
        \ENDFOR
      \ENDIF
    \ENDWHILE
  \end{algorithmic}
\end{algorithm}

%% file: intrinsicOrdering.tex
\subsection{Ordering of the local intrinsic operations}

When optimizing a mesh with respect to both element size and quality, an important factor is the ordering of the local operations. This ordering is chosen to minimize computational cost while maintaining satisfactory results.

The procedure begins by enforcing a Delaunay configuration via edge swaps, ensuring that such a configuration exists. Subsequently, it is beneficial to collapse as many edges as possible, thereby reducing the total number of triangles and edges. Edge splits are then applied, as they are relatively inexpensive, followed by triangle splits on the partially processed mesh. Because each operation is immediately followed by local edge swaps, a global edge swap procedure is only required at the initial stage.

\begin{figure}[t]
  \centering
  \begin{subfigure}{.5\linewidth}
      \centering
      \includegraphics[width=\linewidth]{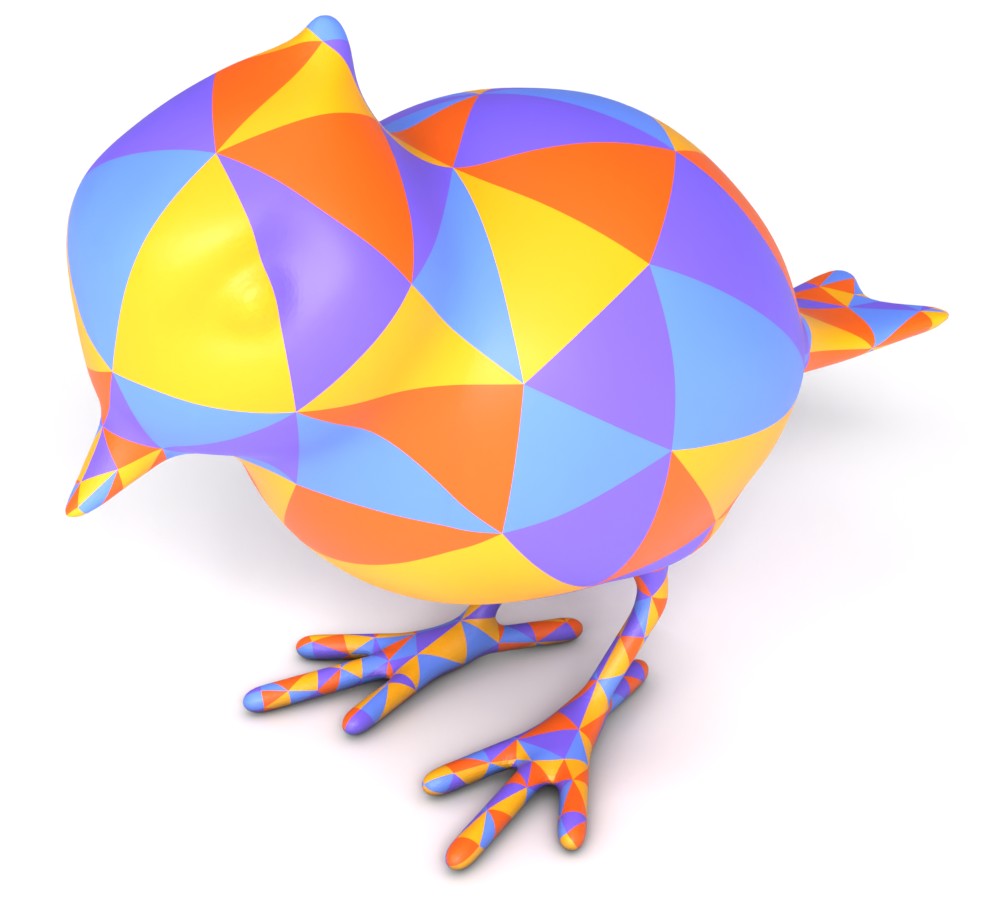}

  \end{subfigure}%

  \hfill%
  \begin{subfigure}{.2\linewidth}
      \centering
      \includegraphics[width=\linewidth]{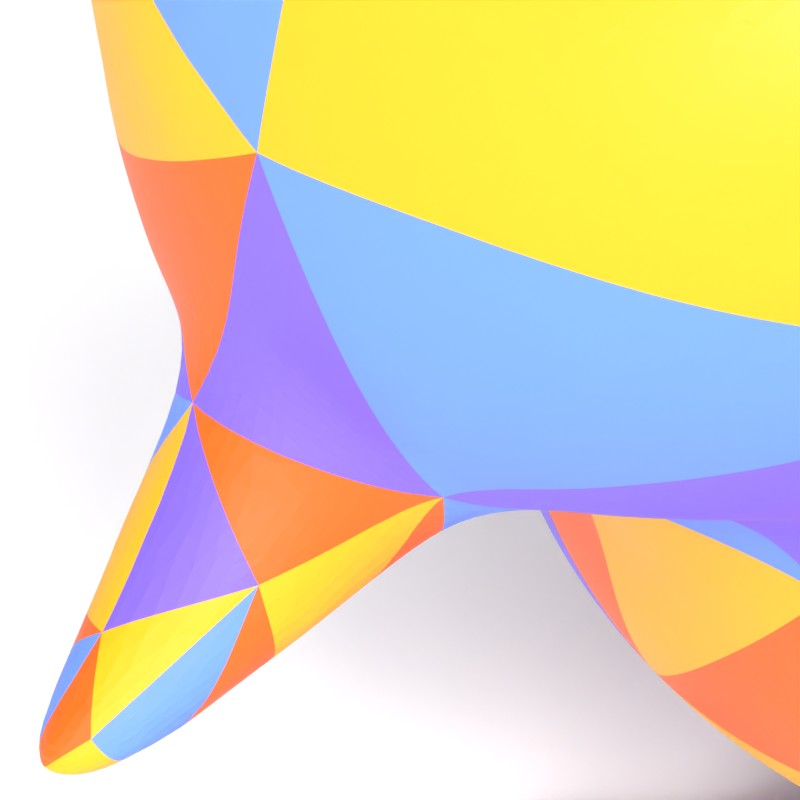}
  \end{subfigure}%
  \hfill%
  \begin{subfigure}{.2\linewidth}
      \centering
      \includegraphics[width=\linewidth]{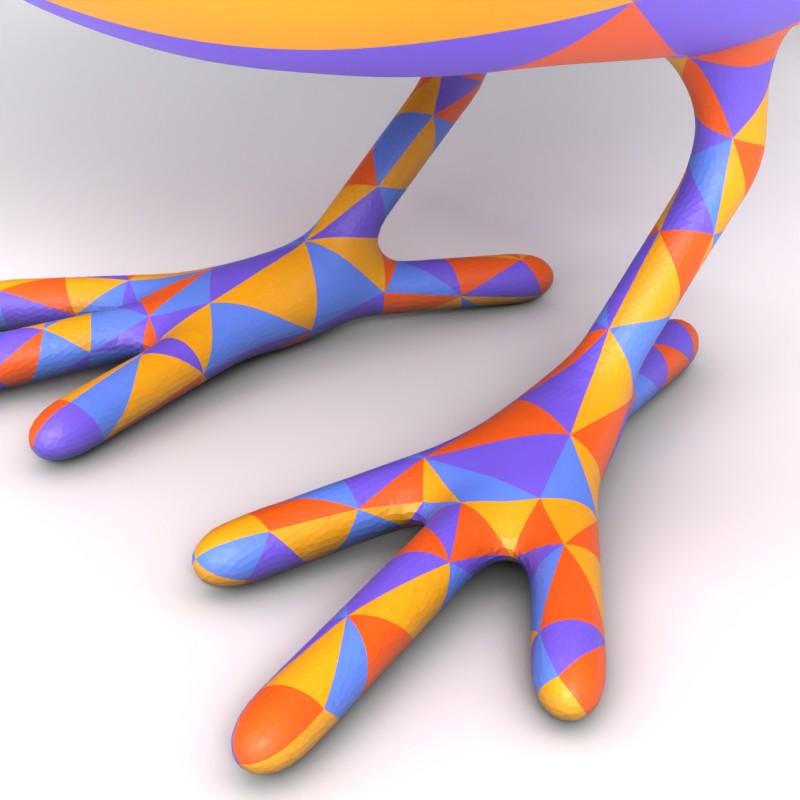}
  \end{subfigure}%
  \hfill%
  \begin{subfigure}{.2\linewidth}
      \centering
      \includegraphics[width=\linewidth]{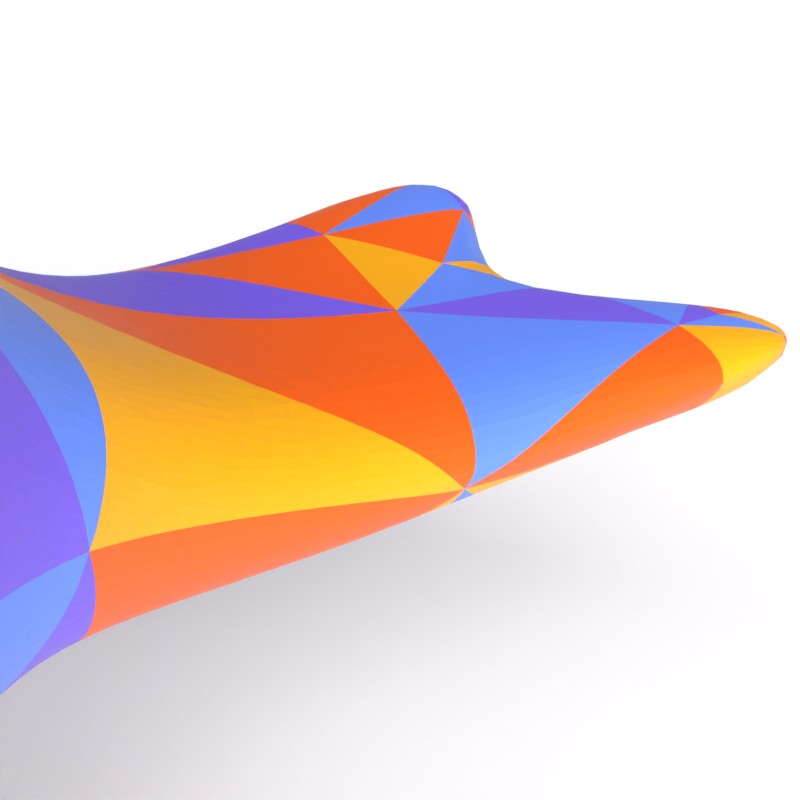}
  \end{subfigure}%
  \hfill%
  \hfill%

  \caption{Example of intrinsic mesh (Thingi10k model ID: 178340).}
  \label{fig:intrinsicMesh}
\end{figure}

This sequence of operations is then iterated until convergence, yielding a mesh that satisfies the prescribed constraints wherever feasible. The complete procedure is summarized in Alg.~\ref{alg:intrinsicMesh} and an example of intrinsic meshing is represented in Fig.~\ref{fig:intrinsicMesh}.

\begin{algorithm}[H]
    \caption{\textsc{intrinsicMesh}()}
    \label{alg:intrinsicMesh}
    \begin{algorithmic}
      \STATE \textsc{swapEdges}(false)
      \STATE \textsc{swapEdges}(true)
        \FOR{$i \leftarrow 1$ \TO $n$}
            \STATE \textsc{collapseEdges}()
            \STATE \textsc{splitEdges}()
            \STATE \textsc{splitTriangles}()
        \ENDFOR
    \end{algorithmic}
\end{algorithm}

%% file: computation.tex
\section{Geodesic Computations}

The intrinsic meshing approach relies on the ability to compute geodesic distances, shortest paths between vertices, and geodesic circumcenters, i.e., points equidistant from three given vertices on the surface. In this work, these computations are performed using continuous Dijkstra methods, specifically the Mitchell-\allowbreak Mount-\allowbreak Papadimitriou (MMP) or Improved Chen-\allowbreak Han (ICH) algorithms. The differences between these two algorithms are briefly discussed in Sec.~\ref{sec:intersections}. To further enhance the efficiency of the method, we propose several minor modifications to these algorithms, specifically designed to reduce the computational cost for the tasks considered here.

\input{computationDistance}

\input{computationBacktrace}

\input{computationAStar}

\input{computationCircumcenter}

%% file: computationDistance.tex
\subsection{Geodesic Distance}

As introduced, continuous Dijkstra algorithms compute exact shortest geodesic distances from given source points on vertices, edges, or faces to any point on a simplicial surface. They resemble the classic graph-based Dijkstra algorithm \cite{dijkstra1959note}, propagating distance information in order of increasing distance. This ensures destinations are reached with minimal distance while avoiding unnecessary propagation. Unlike the graph case, where only scalar values are propagated between vertices, the continuous case requires information sufficient to determine distances to a continuous set of points.

\subsubsection{Triangle Unfolding}

\begin{figure}[b]
  \centering
  \hfill%
  \begin{subfigure}[c]{.3\linewidth}
    \centering
    \resizebox{\linewidth}{!}{\input{figures/SurfacePath.tikz}}
  \end{subfigure}%
  \hfill%
  \begin{subfigure}[c]{.2\linewidth}
    \centering
    \resizebox{\linewidth}{!}{\input{figures/unfolding.tikz}}
  \end{subfigure}%
  \hfill%
  \hfill%
  \caption{Geodesic shortest paths from a point to an interval on the surface and the unfolded faces.}
  \label{fig:unfolding}
\end{figure}
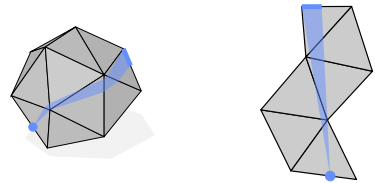

Distance propagation on a triangulated surface occurs along edges connecting adjacent faces, as these edges constitute the only direct links between faces. Furthermore, knowledge of the distance from a source to an edge is sufficient to determine the distance to any point on the surface, since the remaining distance from a point to the edge can be computed within the corresponding face. Consequently, only distances to edges are considered in the following.

By rotating a face around a shared edge, adjacent faces can be aligned within a common plane on opposite sides of the edge, reducing inter-face distance computation to a planar problem. This procedure can be iteratively applied to unfold sequences of faces. However, in such an unfolding, distances from the source to an edge are valid only within a conic region defined by the extremities of the traversed edges, see Fig.~\ref{fig:unfolding}.

Each unfolding therefore defines a continuous set of points on an edge, referred to as an interval or window, which is reachable via a geodesic corresponding to that unfolding. Computing the distance to any point within an interval requires only the relative position of the source in the unfolded plane. Also unfolding an adjacent face corresponds to projecting the interval onto the new edge and rotating the source coordinates in the unfolded plane with respect to this edge. Hence, each unfolding can be fully represented by its associated interval, and the process of computing subsequent unfoldings is equivalent to propagating intervals across neighboring faces.

\subsubsection{Interval Intersections}\label{sec:intersections}

Since the algorithm is applicable to any surface embedded in $\mathbb{R}^3$, regions with nonplanar curvature can give rise to special cases. One such case occurs when two unfoldings generate overlapping intervals along an edge. There are two primary strategies for handling this situation, which distinguish the MMP and CH approaches.

In MMP, the overlap is resolved by computing the point on the edge that is equidistant to the sources of the intersecting intervals. Then the interval sizes are reduced so that they no longer overlap.

In contrast, CH retains overlapping intervals and defers the selection to the distance computation stage, ultimately choosing the interval that yields the shortest path. This strategy leads CH to generate more intervals, as no interval is pruned by another. The improved CH (ICH) approach introduces a filtering mechanism to discard intervals that will not contribute to the shortest path, thereby achieving performance comparable to MMP.

\begin{figure}[t]
  \centering
  \hfill%
  \begin{subfigure}{\linewidth}
    \centering
    \begin{tikzpicture}
        \node[draw=black!20, line width=2, anchor=east] (A) at (0,0) {
          \input{figures/intersection.tikz}
        };
        \node[draw=black!20, line width=2, anchor=west] (B) at (2,+1) {
          \input{figures/intersectionMMP.tikz}
        };
        \draw[-latex, thick, black!20, line width=2] (A) -- (B);
        \node[draw=black!20, line width=2, anchor=west] (C) at (2,-1) {
          \input{figures/intersectionICH.tikz}
        };
        \draw[-latex, thick, black!20, line width=2] (A) -- (C);
    \end{tikzpicture}
  \end{subfigure}%
  \hfill%
  \hfill%
  \caption{Two approaches to handle interval intersections: compute the equidistant point to resolve them (MMP), or retain both intervals and filter those not contributing to the shortest path (ICH).}
  \label{fig:intervalintersections}
\end{figure}
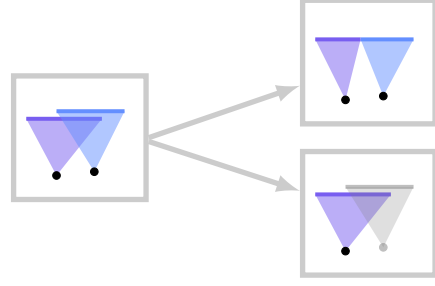

\subsubsection{Geometric Shadow Propagation}

A second special case arises when a window is propagated through a saddle or boundary point. In such situations, certain portions of edges cannot be reached directly, as they lie in a geometric shadow where no direct path exists from the source. Nevertheless, these regions can still be connected via the saddle or boundary point. Consequently, the resulting window is represented with a pseudo-source located at the corresponding position of the saddle or boundary point. Additionally, the window requires an auxiliary parameter $d$ to track the distance from the pseudo-source to the original source, ensuring that the total distance to the source can still be computed efficiently.

\begin{figure}[b]
  \centering
  \hfill%
  \begin{subfigure}{\linewidth}
    \centering
    \begin{tikzpicture}
        \node[draw=black!20, line width=2, anchor=east] (A) at (0,0) {
          \input{figures/shadow.tikz}
        };
        \node[draw=black!20, line width=2, anchor=west] (B) at (1,0) {
          \input{figures/shadowAfter.tikz}
        };
        \draw[-latex, thick, black!20, line width=2] (A) -- (B);
    \end{tikzpicture}
  \end{subfigure}%
  \hfill%
  \hfill%
  \caption{Interval shadow propagation at the boundary and saddle points.}
  \label{fig:intervalintersectionshadow}
\end{figure}
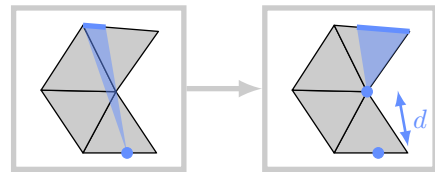

\subsubsection{Propagation Order}

The performance difference between CH and ICH or MMP lies in the order in which intervals are propagated: in ICH and MMP, intervals are processed according to the smallest distance first. This ordering not only enables early termination upon reaching the destination but also facilitates effective pruning of intervals.

Therefore, to implement propagation in this distance-priority order, each window is associated with a parameter $min$, which records the smallest distance from any point of the window to the source. Once all windows with smaller distances have been propagated, the distance to the destination is guaranteed to be minimal.

\subsubsection{Performance Considerations}

In the worst case, the number of windows is $O(n^2)$, as each vertex may split a window into two. Since windows are processed in order of increasing distance, a priority queue is required, leading to an overall complexity of $O(n^2 \log n)$. In practice, however, the number of windows is significantly smaller due to pruning and intersection effects, and propagation is typically restricted to the vicinity of the destination rather than the entire surface, resulting in much lower computational cost.

On top of that, the information associated with each window is relatively lightweight: its corresponding edge $e$, the interval range ($x_{\text{start}}$, $x_{\text{end}}$), the pseudo-source position ($x_{\text{pseudo}}$, $y_{\text{pseudo}}$), the distance $d$ from the pseudo-source to the original source, and the minimal distance value $min$ within the window.

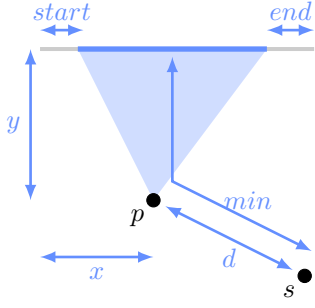
\begin{figure}[t]
  \centering
  \input{figures/interval.tikz}
  \caption{Representation of an interval and all associated information with respect to the source $s$ and the pseudo-source $p$.}
  \label{fig:interval}
\end{figure}

%% file: figures/SurfacePath.tikz
\tdplotsetmaincoords{70}{60}
\begin{tikzpicture}[tdplot_main_coords]
\coordinate (P1) at (-1.1547005383792512, 1.1547005383792512, -1.0);
\coordinate (P2) at (0.0, 0.0, -1.0);
\coordinate (P3) at (0.577350269189626, 0.2886751345948127, -1.0);
\coordinate (P4) at (-2.4424906541753444e-15, 0.8660254037844395, -1.0);
\coordinate (P5) at (0.028446766730999884, -0.22292348979790733, -1.0);
\coordinate (P6) at (0.015068643200552942, -0.4591201405159766, -1.0);
\coordinate (P7) at (-1.3814997262084292, 1.0648803914940785, -1.0);
\coordinate (P8) at (-0.8622846423136789, -0.3930771917559225, -1.0);
\coordinate (P9) at (-2.3869795029440866e-14, 1.3938468501173689, -1.0);
\coordinate (P10) at (-1.2288517677337667, 0.0012860323839596344, -1.0);
\coordinate (P11) at (-1.6794773459897083, 0.5195198703266696, -1.0);
\coordinate (P12) at (-0.9603538343564666, 0.46520251067686036, -1.0);
\coordinate (P13) at (-0.7982722127759704, 0.24246914662684443, -1.0);
\coordinate (P14) at (-0.2673327255909406, 1.1302810914936265, -1.0);
\coordinate (P15) at (-0.8993299239054027, 1.6764469283611128, -1.0);
\coordinate (P16) at (-1.4685139027593284, 1.4829731791077319, -1.0);
\coordinate (P17) at (-0.5268728942554126, 1.5526587011623103, -1.0);

    % \shade[shading=axis, top color=white, bottom color=black!20] (P16) -- (P11) -- (P10) -- (P8) -- (P6) -- (P3) -- (P9)  -- (P17) -- (P15) --  cycle;
    \fill[black!5] (P16) -- (P11) -- (P10) -- (P8) -- (P6) -- (P3) -- (P9)  -- (P17) -- (P15) --  cycle;

    % \node[inner sep=0] at (0,0) {\includegraphics[width=4cm]{sphere.png}};

    \coordinate (1) at (0.00000, 0.00000, 1.00000);
    \coordinate (2) at (0.00000, 0.00000, -1.00000);
    \coordinate (3) at (0.86603, -0.00000, -0.50000);
    \coordinate (4) at (0.86603, -0.00000, 0.50000);
    \coordinate (5) at (0.60314, -0.79762, -0.00460);
    \coordinate (6) at (0.12143, -0.56549, -0.81577);
    \coordinate (7) at (-0.84560, 0.52898, -0.07179);
    \coordinate (8) at (-0.30282, -0.95255, -0.03097);
    \coordinate (9) at (0.57735, 0.81650, 0.00000);
    \coordinate (10) at (-0.85770, -0.36987, -0.35714);
    \coordinate (11) at (-0.88920, -0.27076, 0.36880);
    \coordinate (12) at (0.08479, -0.57994, 0.81024);
    \coordinate (13) at (-0.65399, 0.09819, -0.75010);
    \coordinate (14) at (0.00933, 0.85362, -0.52081);
    \coordinate (15) at (-0.17356, 0.95068, 0.25706);
    \coordinate (16) at (-0.45261, 0.46707, 0.75960);
    \coordinate (17) at (0.44391, 0.58188, 0.68144);
    \definecolor{mycolor}{RGB}{125.57,125.57,125.57}
    \draw[rounded corners=.1pt, fill=mycolor!50!white] (8) -- (6) -- (5) -- cycle;
    \definecolor{mycolor}{RGB}{120.55299999999998,120.55299999999998,120.55299999999998}
    \draw[rounded corners=.1pt, fill=mycolor!50!white] (6) -- (3) -- (5) -- cycle;
    \definecolor{mycolor}{RGB}{98.128,98.128,98.128}
    \draw[rounded corners=.1pt, fill=mycolor!50!white] (11) -- (8) -- (12) -- cycle;
    \definecolor{mycolor}{RGB}{135.74,135.74,135.74}
    \draw[rounded corners=.1pt, fill=mycolor!50!white] (12) -- (8) -- (5) -- cycle;
    \definecolor{mycolor}{RGB}{152.574,152.574,152.574}
    \draw[rounded corners=.1pt, fill=mycolor!50!white] (12) -- (5) -- (4) -- cycle;
    \definecolor{mycolor}{RGB}{127.06499999999998,127.06499999999998,127.06499999999998}
    \draw[rounded corners=.1pt, fill=mycolor!50!white] (4) -- (5) -- (3) -- cycle;
    \definecolor{mycolor}{RGB}{117.90599999999999,117.90599999999999,117.90599999999999}
    \draw[rounded corners=.1pt, fill=mycolor!50!white] (1) -- (12) -- (4) -- cycle;
    \definecolor{mycolor}{RGB}{99.692,99.692,99.692}
    \draw[rounded corners=.1pt, fill=mycolor!50!white] (4) -- (3) -- (9) -- cycle;
    \definecolor{mycolor}{RGB}{101.71199999999999,101.71199999999999,101.71199999999999}
    \draw[rounded corners=.1pt, fill=mycolor!50!white] (1) -- (4) -- (17) -- cycle;
    \definecolor{mycolor}{RGB}{92.008,92.008,92.008}
    \draw[rounded corners=.1pt, fill=mycolor!50!white] (17) -- (4) -- (9) -- cycle;
    \definecolor{mycolor}{RGB}{101.71199999999999,101.71199999999999,101.71199999999999}
    \draw[rounded corners=.1pt, fill=mycolor!50!white] (11) -- (12) -- (1) -- cycle;

    \coordinate (S) at (-0.048265942149835706, -0.7203096898672591, -0.5018505330239273);
    \coordinate (I1) at (0.4698156104054811, -0.8204197636432475, -0.008478439669333714);
    \coordinate (I2) at (0.5500698252815843, -0.775331880993952, 0.0788319547343459);
    \coordinate (I3) at (0.7632604624756995, -0.3118024170962437, 0.3027449772295146);
    \coordinate (I4) at (0.8660254037844373, -2.1211504774498101e-16, 0.4335028522733049);
    \coordinate (I5) at (0.8433061825193056, 0.0642596616794052, 0.46064915446039784);

    \coordinate (I6) at (0.49728517682832474, 0.6757250473627519, 0.4088655653062919);
    \coordinate (I7) at (0.756524293008235, 0.30971591190923, 0.31033851265040624);
    \coordinate (I8) at (0.8660254037844377, -2.121150477449811e-16, 0.22199760414041855);
    
    % \draw (S) -- (I1) -- (I2) -- (I3) -- (I4) -- (I5) -- (17);
    % \draw (I6) -- (I7) -- (I8) -- (5) -- (S);

    \fill[color=myLightBlue, opacity=.5] (S) -- (5) -- (I1) -- cycle;
    \fill[color=myLightBlue, opacity=.5] (I1) -- (5) -- (I2) -- cycle;
    \fill[color=myLightBlue, opacity=.5] (I2) -- (5) -- (I3) -- cycle;
    \fill[color=myLightBlue, opacity=.5] (5) -- (I8) -- (I4) -- (I3) -- cycle;
    \fill[color=myLightBlue, opacity=.5] (I8) -- (I7) -- (I5) -- (I4) -- cycle;
    \fill[color=myLightBlue, opacity=.5] (I7) -- (I6) -- (17) -- (I5) -- cycle;
    
    \fill[myLightBlue] (S) circle (2pt);
    \draw[color=myLightBlue, line width=2] (17) -- (I6);

\end{tikzpicture}

%% file: figures/unfolding.tikz
\begin{tikzpicture}[rotate=-7.5, scale=2]

\definecolor{mycolor}{RGB}{152.574,152.574,152.574}

\newcommand{\drawtriangle}[3]{%
  \draw[rounded corners=.1pt, fill=mycolor!50!white, line width=1] (#1) -- (#2) -- (#3) -- cycle;
}

\coordinate (8) at (0.0, 0.0);
\coordinate (6) at (0.9724762036834154, 0.0);
\coordinate (5) at (0.4356094694401687, 0.8097557696764394);
\drawtriangle{8}{6}{5};
\coordinate (12) at (-0.554140030734521, 0.8304418266021947);
\drawtriangle{8}{5}{12};
\coordinate (4) at (0.0007915517600691029, 1.6877396962729612);
\drawtriangle{12}{5}{4};
\coordinate (3) at (0.9974968734059819, 1.6066317324055248);
\drawtriangle{4}{5}{3};
\coordinate (9) at (0.5693857697414575, 2.5103578429717652);
\drawtriangle{4}{3}{9};
\coordinate (17) at (-0.1569586479789935, 2.412170699734219);
\drawtriangle{4}{9}{17};
\coordinate (1) at (-0.9002972352189837, 2.121374332782332);
% \drawtriangle{4}{17}{1};

\coordinate (S) at ($.4*(8) + .6*(6)$);

\coordinate (I1) at ($1.*(17) + .0*(9)$);
\coordinate (I2) at ($.6*(17) + .4*(9)$);
\coordinate (D) at ($.8*(17) + .2*(9)$);

% \draw[fill=white, fill opacity=.4] (I2) -- (S) -- (I1);
  \fill[color=myLightBlue, opacity=.5] (I2) -- (S) -- (I1);

% \draw[thick, color=myLightBlue] (S) -- (D);
\filldraw[color=myLightBlue] (S) circle (2pt);
% \filldraw (D) circle (1pt);
  \draw[line width=4, color=myLightBlue] (I1) -- (I2);

\end{tikzpicture}

%% file: figures/intersection.tikz
\begin{tikzpicture}
    \useasboundingbox (-0.05,+0.4) rectangle (1.35,-1.);

    % Define coordinates
    \coordinate (A) at (0,-0.05);
    \coordinate (B) at (1,-0.05);
    \coordinate (C) at (0.4,0.05);
    \coordinate (D) at (1.3,0.05);
    \coordinate (P) at (0.4,-.8); % Point above the segment
    \coordinate (Q) at (.9,-.75); % Point above the segment

    \fill[myDarkBlue, opacity=.5] (P) -- (B) -- (A) -- cycle;
    \draw[line width=1.5, color=myDarkBlue] (A) -- (B);
    \fill[myLightBlue, opacity=.5] (Q) -- (D) -- (C) -- cycle;
    \draw[line width=1.5, color=myLightBlue] (C) -- (D);
    %   \filldraw (P) circle (1pt) node[below] {$p$};
    %   \filldraw (Q) circle (1pt) node[below] {$q$};
    \filldraw (P) circle (.05);
    \filldraw (Q) circle (.05);

\end{tikzpicture}

%% file: figures/intersectionMMP.tikz
\begin{tikzpicture}
    \useasboundingbox (-0.05,+0.4) rectangle (1.35,-1.);

    % Define coordinates
    \coordinate (A) at (0,0);
    \coordinate (B) at (1,0);
    \coordinate (C) at (0.6,0);
    \coordinate (D) at (1.3,0);
    \coordinate (P) at (0.4,-.8); % Point above the segment
    \coordinate (Q) at (.9,-.75); % Point above the segment

    \fill[myDarkBlue, opacity=.5] (P) -- (C) -- (A) -- cycle;
    \draw[line width=1.5, color=myDarkBlue] (A) -- (C);
    \fill[myLightBlue, opacity=.5] (Q) -- (D) -- (C) -- cycle;
    \draw[line width=1.5, color=myLightBlue] (C) -- (D);
    %   \filldraw (P) circle (1pt) node[below] {$p$};
    %   \filldraw (Q) circle (1pt) node[below] {$q$};
    \filldraw (P) circle (.05);
    \filldraw (Q) circle (.05);

\end{tikzpicture}

%% file: figures/intersectionICH.tikz
\begin{tikzpicture}
    \useasboundingbox (-0.05,+0.4) rectangle (1.35,-1.);

    % Define coordinates
    \coordinate (A) at (0,-0.05);
    \coordinate (B) at (1,-0.05);
    \coordinate (C) at (0.4,0.05);
    \coordinate (D) at (1.3,0.05);
    \coordinate (P) at (0.4,-.8); % Point above the segment
    \coordinate (Q) at (.9,-.75); % Point above the segment

    \fill[myDarkBlue, opacity=.5] (P) -- (B) -- (A) -- cycle;
    \draw[line width=1.5, color=myDarkBlue] (A) -- (B);
    \fill[gray, opacity=.25] (Q) -- (D) -- (C) -- cycle;
    \draw[line width=1.5, color=gray, opacity=.5] (C) -- (D);
    %   \filldraw (P) circle (1pt) node[below] {$p$};
    %   \filldraw (Q) circle (1pt) node[below] {$q$};
    \filldraw (P) circle (.05);
    \filldraw[opacity=.5, gray] (Q) circle (.05);

\end{tikzpicture}

%% file: figures/shadow.tikz
\begin{tikzpicture}
    \useasboundingbox (-.8,-0.1) rectangle (1.1,1.8);
    \definecolor{mycolor}{RGB}{152.574,152.574,152.574}

    \newcommand{\drawtriangle}[3]{%
    \draw[rounded corners=.1pt, fill=mycolor!50!white, line width=.5] (#1) -- (#2) -- (#3) -- cycle;
    }

    \coordinate (8) at (0.0, 0.0);
    \coordinate (6) at (0.9724762036834154, 0.0);
    \coordinate (5) at (0.4356094694401687, 0.8097557696764394);
    \drawtriangle{8}{6}{5};
    \coordinate (12) at (-0.554140030734521, 0.8304418266021947);
    \drawtriangle{8}{5}{12};
    \coordinate (4) at (0.0007915517600691029, 1.6877396962729612);
    \drawtriangle{12}{5}{4};
    \coordinate (3) at (0.9974968734059819, 1.6066317324055248);
    \drawtriangle{4}{5}{3};

    \coordinate (S) at ($.4*(8) + .6*(6)$);
    \coordinate (I1) at ($1.*(4) + .0*(3)$);
    \coordinate (I2) at ($.7*(4) + .3*(3)$);

    \fill[color=myLightBlue, opacity=.5] (I2) -- (S) -- (I1);
    \filldraw[color=myLightBlue] (S) circle (2pt);
    \draw[line width=2, color=myLightBlue] (I1) -- (I2);

\end{tikzpicture}

%% file: figures/shadowAfter.tikz
\begin{tikzpicture}
    \useasboundingbox (-.8,-0.1) rectangle (1.1,1.8);
\definecolor{mycolor}{RGB}{152.574,152.574,152.574}

    \newcommand{\drawtriangle}[3]{%
    \draw[rounded corners=.1pt, fill=mycolor!50!white, line width=.5] (#1) -- (#2) -- (#3) -- cycle;
    }

    \coordinate (8) at (0.0, 0.0);
    \coordinate (6) at (0.9724762036834154, 0.0);
    \coordinate (5) at (0.4356094694401687, 0.8097557696764394);
    \drawtriangle{8}{6}{5};
    \coordinate (12) at (-0.554140030734521, 0.8304418266021947);
    \drawtriangle{8}{5}{12};
    \coordinate (4) at (0.0007915517600691029, 1.6877396962729612);
    \drawtriangle{12}{5}{4};
    \coordinate (3) at (0.9974968734059819, 1.6066317324055248);
    \drawtriangle{4}{5}{3};

    \coordinate (S) at ($.4*(8) + .6*(6)$);
    \coordinate (I1) at ($1.*(4) + .0*(3)$);
    \coordinate (I2) at ($.7*(4) + .3*(3)$);

    \fill[color=myLightBlue, opacity=.5] (I2) -- (5) -- (3);
    \filldraw[myLightBlue] (S) circle (2pt);
    \filldraw[myLightBlue] (5) circle (2pt);
    \draw[line width=2, color=myLightBlue] (3) -- (I2);

    \draw[latex-latex, myLightBlue, line width=1, shorten <=1, shorten >=1] ($(S)+(0.4,0.05)$) -- ($(5)+(0.4,0.05)$) node[midway, right] {$d$};

\end{tikzpicture}

%% file: figures/interval.tikz
\newcommand{\dx}{0.4}
\newcommand{\dy}{-.8}
\begin{tikzpicture}[scale=2.5]

  % Define coordinates
  \coordinate (A) at (0,0);
  \coordinate (B) at (1,0);
  \coordinate (start) at (-0.2,0);
  \coordinate (end) at (1.25,0);
  \coordinate (P) at (\dx,\dy); % Point above the segment
  \coordinate (S) at (1.2,-1.2); % True source

  \fill[myLightBlue, opacity=.3] (P) -- (B) -- (A) -- cycle;
  \draw[line width=1.5, color=black!20] (start) -- (end);
  \draw[line width=2, color=myLightBlue] (A) -- (B);
  \filldraw (P) circle (1pt) node[below left] {$p$};
  \filldraw (S) circle (1pt) node[below left] {$s$};

  \draw[latex-latex, myLightBlue, line width=1] (start) ++(0,0.1) -- ++(.23,0.) node[midway, above] {$start$};
  \draw[latex-latex, myLightBlue, line width=1] (B) ++(0,0.1) -- ++(0.25,0) node[midway, above] {$end$};
  \draw[latex-latex, myLightBlue, line width=1] (A) ++(-.2,\dy-0.3) -- ++(\dx+.2,0) node[midway, below] {$x$};
  \draw[latex-latex, myLightBlue, line width=1] (A) ++(-0.25,0) -- ++(0,\dy) node[midway, left] {$y$};
  \draw[latex-latex, line width=1, myLightBlue, shorten <=5, shorten >=5] (P) -- (S) node[midway, below] {$d$};
  \draw[latex-latex, line width=1, myLightBlue, shorten <=10, shorten >=5] ($(\dx,0)+(0.1,0.1)$) -- ($(P)+(0.1,0.1)$) -- ($(S)+(0.1,0.1)$) node[midway, above] {$min$};
\end{tikzpicture}

%% file: computationBacktrace.tex
\subsection{Geodesic Shortest Path}\label{sec:backtracking}

The shortest path is recovered by backtracking from the destination to the source, following windows across edges that minimize the distance. This process identifies a sequence of points on edges and vertices forming the path, and consists of iterating over candidate edges and selecting the shortest path through the corresponding windows.

\subsubsection{Candidate Edges}

The set of candidate windows depends on the type of surface point considered. For a face, distances are evaluated to windows on its edges. For an edge, they are evaluated on the opposite edges of adjacent faces. At a vertex, only opposite edges are considered, as adjacent edges would introduce ambiguity due to equal distances. The same procedure applies to MMP and ICH, with the difference that more windows are present per edge in ICH since windows are not intersected. However, since intervals remain unchanged in ICH, one can store additional information identifying their origin interval, similarly to Dijkstra’s algorithm on graphs. The backtrace then reduces to following the chain of ancestor intervals.

\subsubsection{Distance through a window}\label{sec:distancethroughwindow}

The distance from a point to the source through a window is computed in the unfolded plane, where the pseudo-source is reflected across the edge. The minimal distance is then given either by the direct path through the window or via its endpoints.

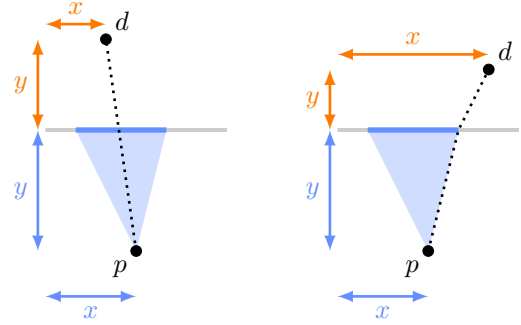
\begin{figure}[htbp]
  \centering
  \begin{subfigure}{.5\linewidth}
    \centering
    \input{figures/throughWindow.tikz}
  \end{subfigure}%
  \begin{subfigure}{.5\linewidth}
    \centering
    \input{figures/throughWindowBis.tikz}
  \end{subfigure}
  \caption{Representation of candidate edges and shortest paths through a window.}
  \label{fig:backtrace}
\end{figure}

\subsubsection{Performance Considerations}

The backtracing cost is proportional to the number of windows, bounded by $O(n)$, and more importantly to the number of points $k$ in the path. Its worst-case complexity is $O(nk)$, which remains negligible compared to the cost of distance propagation. Consequently, computational efficiency must primarily be achieved during the propagation stage.

%% file: figures/throughWindow.tikz
\newcommand{\dx}{0.4}
\newcommand{\dy}{-.8}
\begin{tikzpicture}[scale=2]

  % Define coordinates
  \coordinate (A) at (0,0);
  \coordinate (B) at (.6,0);
  \coordinate (start) at (-0.2,0);
  \coordinate (end) at (1,0);
  \coordinate (P) at (\dx,\dy); % Point above the segment
  \coordinate (S) at (1.2,-1.2); % True source
  \coordinate (D) at (0.2,+0.6);

  \fill[myLightBlue, opacity=.3] (P) -- (B) -- (A) -- cycle;
  \draw[line width=1.5, color=black!20] (start) -- (end);
  \draw[line width=2, color=myLightBlue] (A) -- (B);
  \filldraw (P) circle (1pt) node[below left] {$p$};
  \filldraw (D) circle (1pt) node[above right] {$d$};
  % \filldraw (S) circle (1pt) node[below left] {$s$};

  % \draw[latex-latex, myLightBlue, line width=1] (start) ++(0,0.1) -- ++(.23,0.) node[midway, above] {start};
  % \draw[latex-latex, myLightBlue, line width=1] (B) ++(0,0.1) -- ++(0.25,0) node[midway, above] {end};
  \draw[latex-latex, myLightBlue, line width=1] (A) ++(-.2,\dy-0.3) -- ++(\dx+.2,0) node[midway, below] {$x$};
  \draw[latex-latex, myLightBlue, line width=1] (A) ++(-0.25,0) -- ++(0,\dy) node[midway, left] {$y$};
  % \draw[latex-latex, line width=1, myLightBlue, shorten <=5, shorten >=5] (P) -- (S) node[midway, below] {$d$};
  % \draw[latex-latex, line width=1, myLightBlue, shorten <=10, shorten >=5] ($(\dx,0)+(0.1,0.1)$) -- ($(P)+(0.1,0.1)$) -- ($(S)+(0.1,0.1)$) node[midway, above] {$min$};
  \draw[dotted, line width=1, black] (P) -- (D);

  \draw[latex-latex, mycolorYO, line width=1] (A) ++(-.2,.7) -- ++(.4,0) node[midway, above] {$x$};
  \draw[latex-latex, mycolorYO, line width=1] (A) ++(-0.25,0.) -- ++(0,.6) node[midway, left] {$y$};
\end{tikzpicture}

%% file: figures/throughWindowBis.tikz
\newcommand{\dx}{0.4}
\newcommand{\dy}{-.8}
\begin{tikzpicture}[scale=2]

  % Define coordinates
  \coordinate (A) at (0,0);
  \coordinate (B) at (.6,0);
  \coordinate (start) at (-0.2,0);
  \coordinate (end) at (1,0);
  \coordinate (P) at (\dx,\dy); % Point above the segment
  \coordinate (S) at (1.2,-1.2); % True source
  \coordinate (D) at (0.8,+0.4);

  \fill[myLightBlue, opacity=.3] (P) -- (B) -- (A) -- cycle;
  \draw[line width=1.5, color=black!20] (start) -- (end);
  \draw[line width=2, color=myLightBlue] (A) -- (B);
  \filldraw (P) circle (1pt) node[below left] {$p$};
  \filldraw (D) circle (1pt) node[above right] {$d$};
  % \filldraw (S) circle (1pt) node[below left] {$s$};

  % \draw[latex-latex, myLightBlue, line width=1] (start) ++(0,0.1) -- ++(.23,0.) node[midway, above] {start};
  % \draw[latex-latex, myLightBlue, line width=1] (B) ++(0,0.1) -- ++(0.25,0) node[midway, above] {end};
  \draw[latex-latex, myLightBlue, line width=1] (A) ++(-.2,\dy-0.3) -- ++(\dx+.2,0) node[midway, below] {$x$};
  \draw[latex-latex, myLightBlue, line width=1] (A) ++(-0.25,0) -- ++(0,\dy) node[midway, left] {$y$};
  % \draw[latex-latex, line width=1, myLightBlue, shorten <=5, shorten >=5] (P) -- (S) node[midway, below] {$d$};
  % \draw[latex-latex, line width=1, myLightBlue, shorten <=10, shorten >=5] ($(\dx,0)+(0.1,0.1)$) -- ($(P)+(0.1,0.1)$) -- ($(S)+(0.1,0.1)$) node[midway, above] {$min$};
  \draw[dotted, line width=1, black] (P) -- (B) -- (D);

  \draw[latex-latex, mycolorYO, line width=1] (A) ++(-.2,.5) -- ++(1.,0) node[midway, above] {$x$};
  \draw[latex-latex, mycolorYO, line width=1] (A) ++(-0.25,0.) -- ++(0,.4) node[midway, left] {$y$};
\end{tikzpicture}

%% file: computationAStar.tex
\subsection{Geodesic A* Search}\label{sec:astar}

Introduced in \cite{hart1968formal}, the A* search algorithm improves upon Dijkstra’s method on graphs by incorporating information about the goal’s location to guide the search. While Dijkstra computes shortest paths to all nodes, A* focuses on a single target, which aligns with our needs. The algorithm prioritizes nodes based on an estimate of the total path length given by the exact distance from the source and a heuristic lower bound on the distance to the destination. Consequently, A* applies to graphs where such admissible heuristics exist.

In our setting, an admissible heuristic is readily available, as the Euclidean distance between two points provides a lower bound on their geodesic distance. Accordingly, the window minimum distance can be modified to reflect the estimated total distance from source to destination through the window. This estimate is obtained by reusing the point-to-source distance computation introduced for backtracing, see Sec.~\ref{sec:distancethroughwindow}. This adjustment is sufficient to guide the propagation, as the priority queue relies on these minimal values to select windows. Moreover, after the destination is first reached, propagation must continue for windows whose estimated total distance remains smaller than the current best solution, ensuring optimality as in the classical MMP and ICH algorithms.

\begin{figure}
  \centering
  \hfill%
  \begin{subfigure}{.4\linewidth}
      \includegraphics[width=\linewidth]{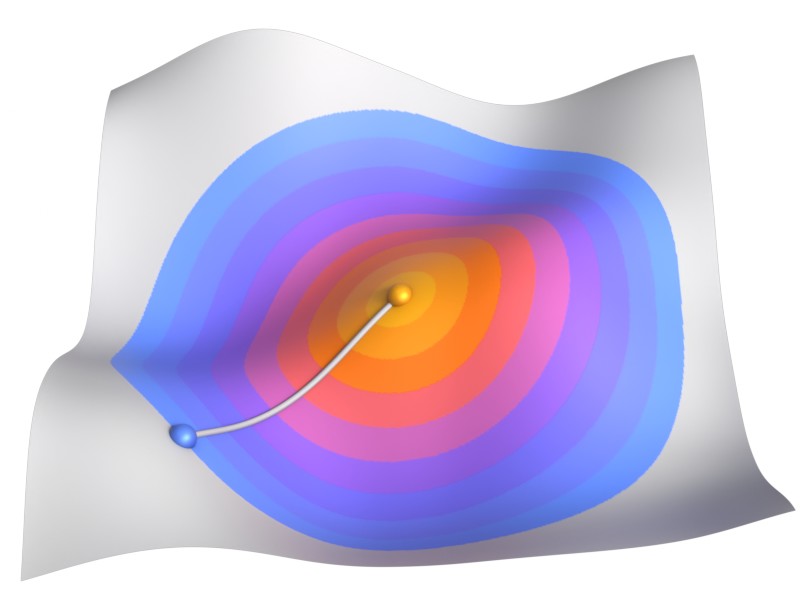}
  \end{subfigure}%
  \hfill%
  \begin{subfigure}{.4\linewidth}
      \includegraphics[width=\linewidth]{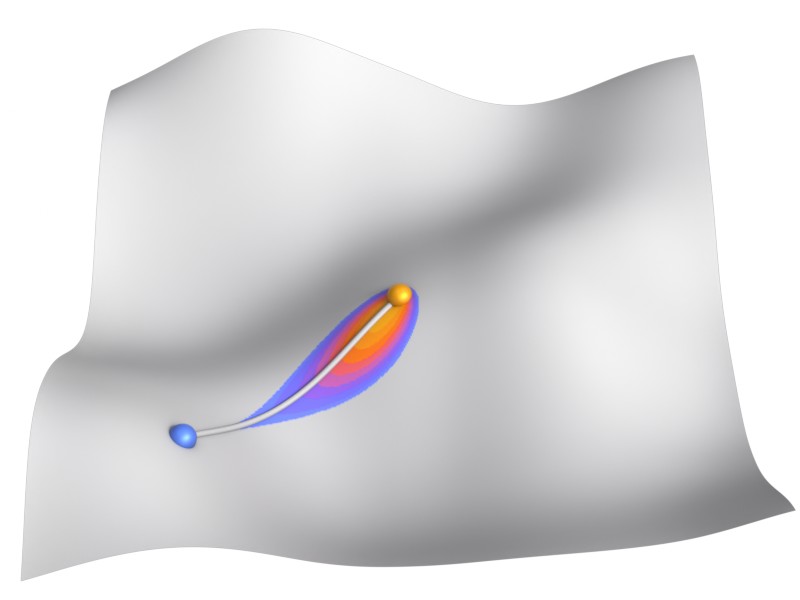}
  \end{subfigure}%
  \hfill%
  \hfill%
  \caption{Geodesic shortest path computed by sorting intervals with respect to the distance to the source (left) and the A* heuristic distance to the destination (right).}
  \label{fig:astar}
\end{figure}

Fig.~\ref{fig:astar} compares the A* search with a classical continuous Dijkstra algorithm. The A* method propagates distances selectively toward the destination, reducing the number of windows to only $5\%$ of the original. This reduction occurs because edges far from the source generate many intervals due to intermediate vertices, most of which are avoided with the guided search. In addition to fewer windows, the priority queue size is smaller, as fewer windows are active at the propagation front. Overall, computing geodesics on a model with a thousand faces requires only $3\%$ of the time compared to the classical MMP algorithm. This speedup is critical for enabling the intrinsic meshing approach presented in this work.

%% file: computationCircumcenter.tex
\subsection{Geodesic Circumcenter}\label{sec:circumcenter}

A fundamental component of Delaunay refinement is the circumcenter of a triangle. As discussed in Sec.~\ref{sec:existandunique}, the existence and uniqueness of the circumcenter are not guaranteed for arbitrary configurations. However, in the following, such a circumcenter is assumed to exist and be unique.

For surfaces embedded in three-dimensional space, the notion of a circumcircle generalizes naturally as the locus of points equidistant from the triangle’s vertices. This formulation allows geodesic distances to be used to define and compute the circumcenter.

A method for computing circumcenters was proposed in \cite{liu2010construction}, where triangles containing the circumcenter are recursively refined using intervals along the edges until the circumcenter is captured by an edge. However, since only circumcenters are required, rather than the full Voronoi diagram, a simplified strategy is adopted here.

Again, distances are propagated from the three vertices, and candidate triangles potentially containing the circumcenter are identified using edge intervals. Unlike the previous approach, all possible circumcenter positions arising from combinations of edge windows are evaluated. Each candidate is then verified to ensure it lies within the triangle and satisfies the equidistance condition. The only requirement for this method is the ability to compute a circumcenter from three distance windows.

\begin{figure}[htbp]
  \centering
  \input{figures/circumcenter.tikz}
  \caption{Representation of the circumcenter as the intersection of geodesic bisectors for three intervals with the three associated pseudo-sources.}
  \label{fig:circumcenter}
\end{figure}
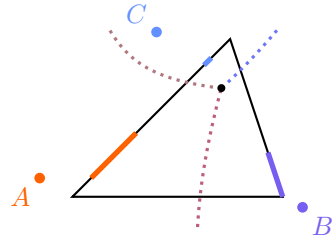

In the planar case, circumcenters are obtained by intersecting edge bisectors, which are straight lines. On curved surfaces, bisectors are generally non-linear: points equidistant from two pseudo-sources satisfy a hyperbolic equation as represented in Fig.~\ref{fig:circumcenter}. Nevertheless, an analytical solution for the intersection of such bisectors can be derived, as detailed in Appendix~\ref{sec:circumcenterformula}.

To improve the propagation speed of distance intervals toward the circumcenter, a strategy similar to that presented in Section~\ref{sec:astar} can be employed. In this context, however, the final position of the circumcenter is unknown and constitutes the quantity to be determined. As an estimate, one may use the center of the smallest sphere passing through the three points as a target. This heuristic performs well in nearly planar settings, but it becomes unreliable in the presence of strong surface curvature, thereby limiting the efficiency of this optimization.

%% file: figures/circumcenter.tikz
\begin{tikzpicture}[x=0.1\linewidth, y=0.1\linewidth]

\def\x{1.8}
\def\y{1.8}

\coordinate (V0) at (-0.8*\x, 0.1*\y);
\coordinate (V1) at (+1.2*\x, 0.1*\y);
\coordinate (V2) at (+0.7*\x, 1.6*\y);

\draw[thick] (V0) -- (V1) -- (V2) -- cycle;
\draw[line width=2, myOrange, shorten <=10, shorten >=50] (V0) -- (V2);
\draw[line width=2, myLightBlue, shorten <=70, shorten >=10] (V0) -- (V2);
\draw[line width=2, myDarkBlue, shorten <=0, shorten >=45] (V1) -- (V2);

% Points
\def\ax{-2.0}
\def\ay{+0.5}
\def\bx{+2.5}
\def\by{+0.0}
\def\cx{+0.0}
\def\cy{+3.0}
\coordinate (A) at (\ax,\ay);
\coordinate (B) at (\bx,\by);
\coordinate (C) at (\cx,\cy);

\fill[myOrange] (A) circle (2pt) node[below left, myOrange] {$A$};
\fill[myDarkBlue] (B) circle (2pt) node[below right, myDarkBlue] {$B$};
\fill[myLightBlue] (C) circle (2pt) node[above left, myLightBlue] {$C$};

% Parameters
\def\a{.5}      % hyperbola parameter
\def\c{sqrt((\bx-\ax)^2+(\by-\ay)^2)/2}      % half distance between foci
\pgfmathsetmacro{\b}{sqrt(\c*\c - \a*\a)}

% --- Hyperbola with foci A,B ---
\begin{scope}
  \pgfmathsetmacro{\angleAB}{atan2(\by - \ay,\bx - \ax)}
  \pgfmathsetmacro{\midABx}{\ax + (\bx - \ax)/2}
  \pgfmathsetmacro{\midABy}{\ay + (\by - \ay)/2}
  \begin{scope}[shift={({\midABx},{\midABy})}, rotate=\angleAB]
    \draw[dotted, very thick, domain=-.24:.775, samples=20, color=myOrange!50!myDarkBlue] 
      plot ({ \a*cosh(\x) },{ \b*sinh(\x) });
  \end{scope}
\end{scope}
% \draw[blue, thick, domain=-1:1, samples=100]
%   plot ({ -\a*cosh(\x) },{ \b*sinh(\x) });

\def\a{1.}      % hyperbola parameter
\def\c{sqrt((\cx-\ax)^2+(\cy-\ay)^2)/2}      % half distance between foci
\pgfmathsetmacro{\b}{sqrt(\c*\c - \a*\a)}

% --- Hyperbola with foci A,C ---
\begin{scope}
  \pgfmathsetmacro{\angleAC}{atan2(\cy - \ay,\cx - \ax)}
  \pgfmathsetmacro{\midACx}{\ax + (\cx - \ax)/2}
  \pgfmathsetmacro{\midACy}{\ay + (\cy - \ay)/2}

  \begin{scope}[shift={({\midACx},{\midACy})}, rotate=\angleAC]
    \draw[dotted, very thick, domain=-1.:.52, samples=20, color=myOrange!50!myLightBlue]
      plot ({ \a*cosh(\x) },{ \b*sinh(\x) });
    % \draw[red, thick, domain=-1:1, samples=100]
    %   plot ({ -\a*cosh(\x) },{ \b*sinh(\x) });
  \end{scope}
\end{scope}

\def\a{0.5}      % hyperbola parameter
\def\c{sqrt((\cx-\bx)^2+(\cy-\by)^2)/2}      % half distance between foci
\pgfmathsetmacro{\b}{sqrt(\c*\c - \a*\a)}

% --- Hyperbola with foci B,C ---
\begin{scope}
  \pgfmathsetmacro{\angleBC}{atan2(\cy - \by,\cx - \bx)}
  \pgfmathsetmacro{\midBCx}{\bx + (\cx - \bx)/2}
  \pgfmathsetmacro{\midBCy}{\by + (\cy - \by)/2}

  \begin{scope}[shift={({\midBCx},{\midBCy})}, rotate=\angleBC]
    \draw[dotted, very thick, domain=-.77:-.1, samples=20, color=myDarkBlue!50!myLightBlue]
      plot ({ \a*cosh(\x) },{ \b*sinh(\x) });
    % \draw[green!70!black, thick, domain=-.5:1.5, samples=100]
    %   plot ({ -\a*cosh(\x) },{ \b*sinh(\x) });
  \end{scope}
\end{scope}

\fill[black] (1.11, 2.04) circle (1.5pt);

\end{tikzpicture}

%% file: result.tex
\section{Experimental Results}

\subsection{Effect of constraints}

Fig.~\ref{fig:examples} illustrates various test cases for intrinsic remeshing under varying angular and minimum characteristic lengths constraints. A primary observation is that these intrinsic elements, even if topologically defined as triangles, may exhibit complex geometries. Because they are defined as a subdivision of the underlying input manifold, a single intrinsic triangle often spans a collection of original mesh faces leading to discontinuous changes in its shape. However, as the sampling density increases, the geometric complexity of these intrinsic triangles decreases. In this high-density regime, the elements converge toward simpler shapes.

Furthermore, the results demonstrate that imposing strict angular bounds effectively drives adaptive refinement. In regions of high geometric curvature, typically associated with critical simulation phenomena, fitting large intrinsic triangles that satisfy prescribed angular limits is difficult. Consequently, the algorithm naturally produces a denser triangulation in these zones to satisfy the constraints.

Finally, while the prescribed angular bounds are not guaranteed due to the numerical challenges of locating intrinsic circumcenters and avoiding intersecting geodesics on highly curved surfaces, the majority of elements successfully adhere to the constraints. It should be noted, however, that strictly enforcing these bounds significantly increases the total element count.

\newcommand{\width}{0.25\linewidth}
\begin{figure*}
  \centering
  \hfill%
  \begin{subfigure}{\width}
      \centering
      \includegraphics[width=\linewidth]{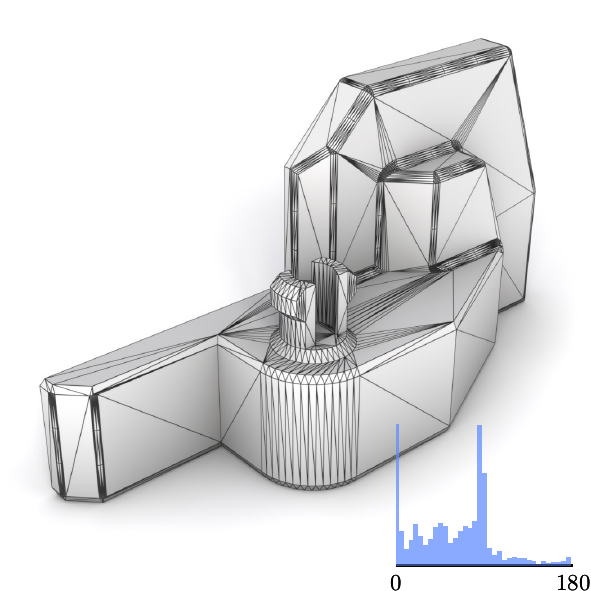}
      \caption{Initial mesh (\#f=2,074)}
  \end{subfigure}%
  \hfill%
  \begin{subfigure}{\width}
      \centering
      \includegraphics[width=\linewidth]{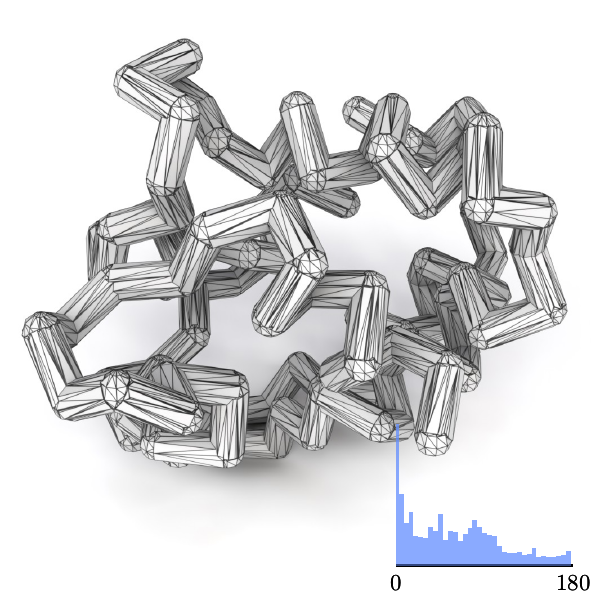}
      \caption{Initial mesh (\#f=20,180)}
  \end{subfigure}%
  \hfill%
  \begin{subfigure}{\width}
      \centering
      \includegraphics[width=\linewidth]{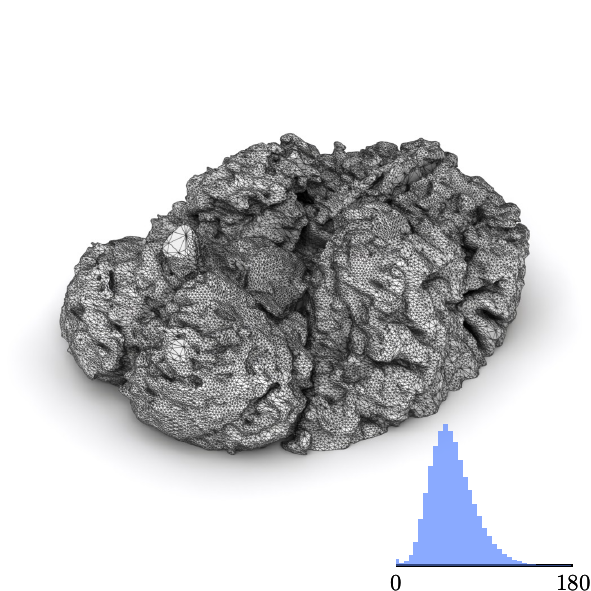}
      \caption{Initial mesh (\#f=200,192)}
  \end{subfigure}%
  \hfill%
  \hfill%

  \hfill%
  \begin{subfigure}{\width}
      \centering
      \includegraphics[width=\linewidth]{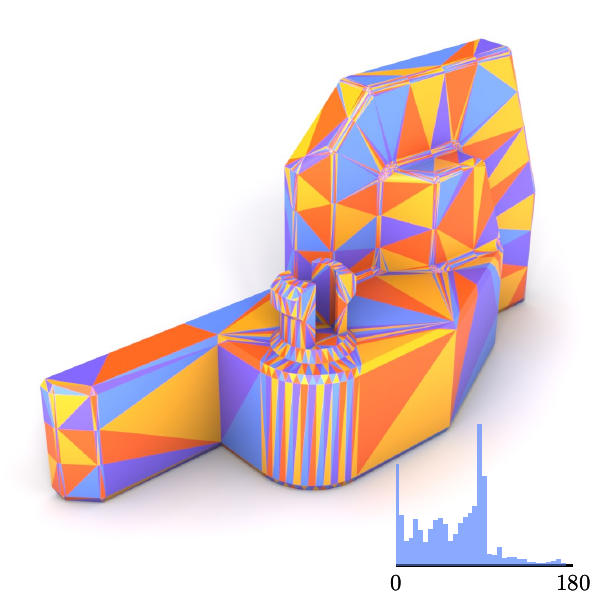}
      \caption{Delaunay triangulation (\#f=2,082)}
  \end{subfigure}%
  \hfill%
  \begin{subfigure}{\width}
      \centering
      \includegraphics[width=\linewidth]{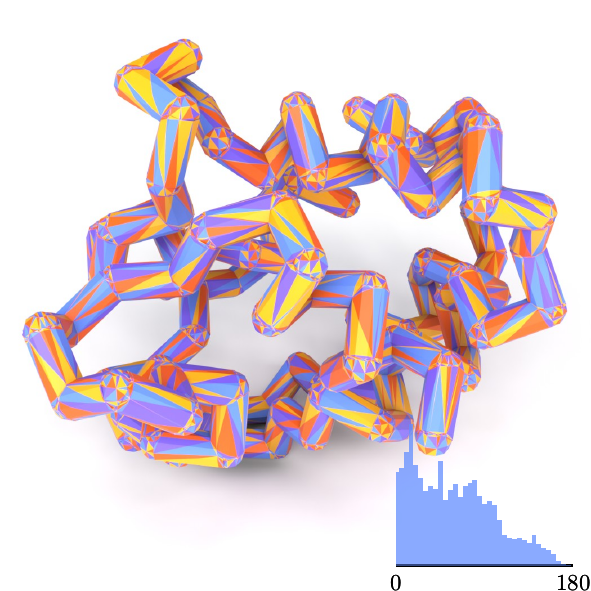}
      \caption{Delaunay triangulation (\#f=20,264)}
  \end{subfigure}%
  \hfill%
  \begin{subfigure}{\width}
      \centering
      \includegraphics[width=\linewidth]{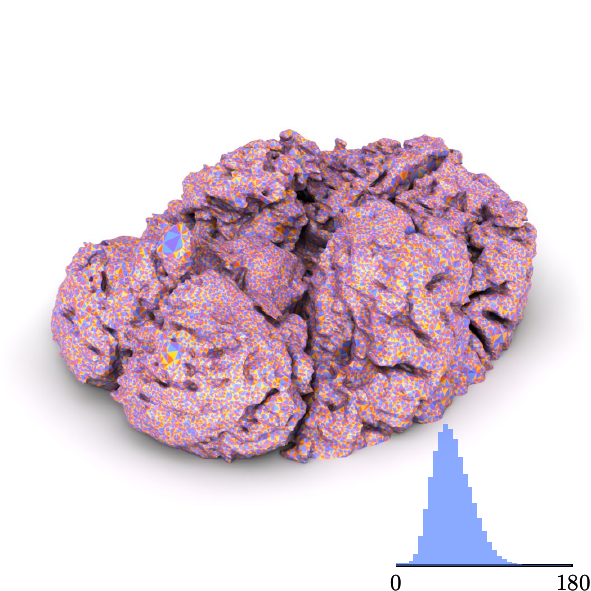}
      \caption{Delaunay triangulation (\#f=200,434)}
  \end{subfigure}%
  \hfill%
  \hfill%

  \hfill%
  \begin{subfigure}{\width}
      \centering
      \includegraphics[width=\linewidth]{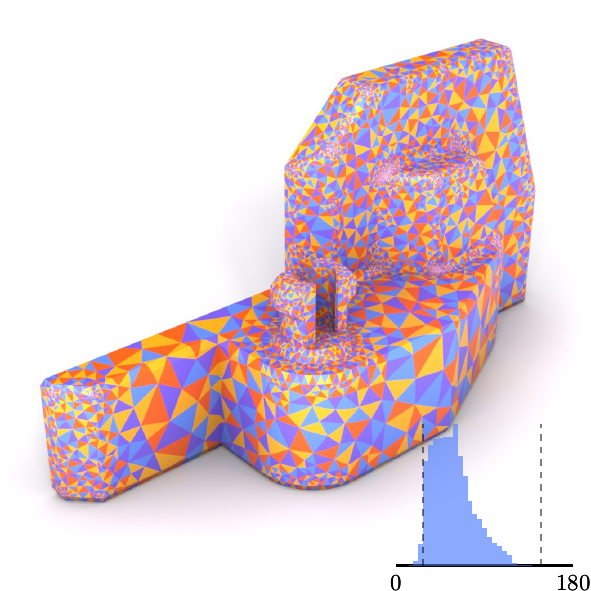}
      \caption{$30^\circ < \theta < 120^\circ$ (\#f=10,434)}
  \end{subfigure}%
  \hfill%
  \begin{subfigure}{\width}
      \centering
      \includegraphics[width=\linewidth]{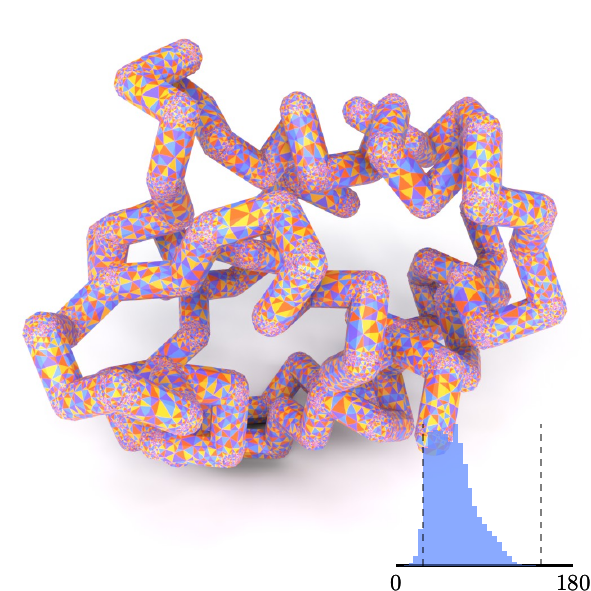}
      \caption{$30^\circ < \theta < 120^\circ$ (\#f=122,662)}
  \end{subfigure}%
  \hfill%
  \begin{subfigure}{\width}
      \centering
      \includegraphics[width=\linewidth]{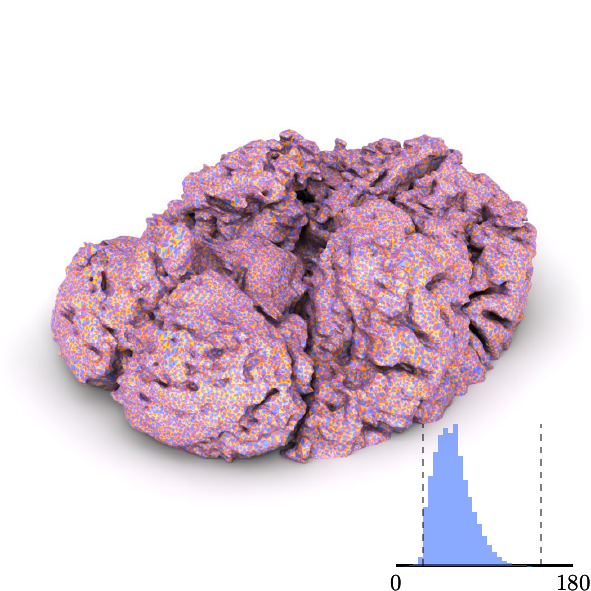}
      \caption{$30^\circ < \theta < 120^\circ$ (\#f=259,646)}
  \end{subfigure}%
  \hfill%
  \hfill%

  \hfill%
  \begin{subfigure}{\width}
      \centering
      \includegraphics[width=\linewidth]{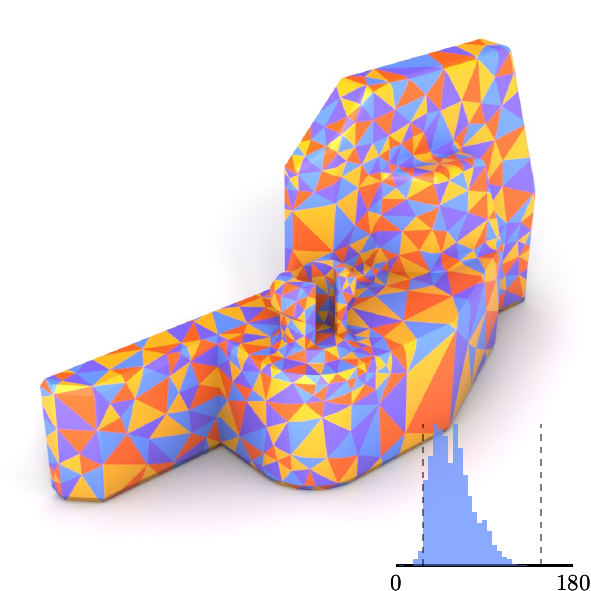}
      \caption{$30^\circ < \theta < 120^\circ$, $cl_{\text{min}} = 1$ (\#f=1,168)}
  \end{subfigure}%
  \hfill%
  \begin{subfigure}{\width}
      \centering
      \includegraphics[width=\linewidth]{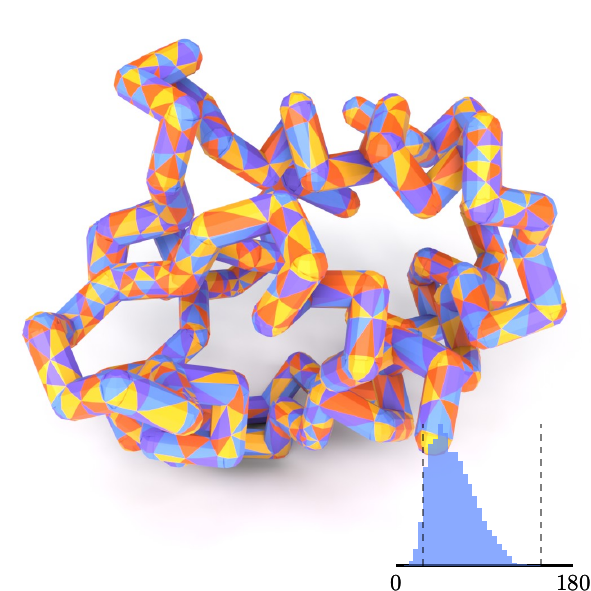}
      \caption{$30^\circ < \theta < 120^\circ$, $cl_{\text{min}} = 1$ (\#f=2,340)}
  \end{subfigure}%
  \hfill%
  \begin{subfigure}{\width}
      \centering
      \includegraphics[width=\linewidth]{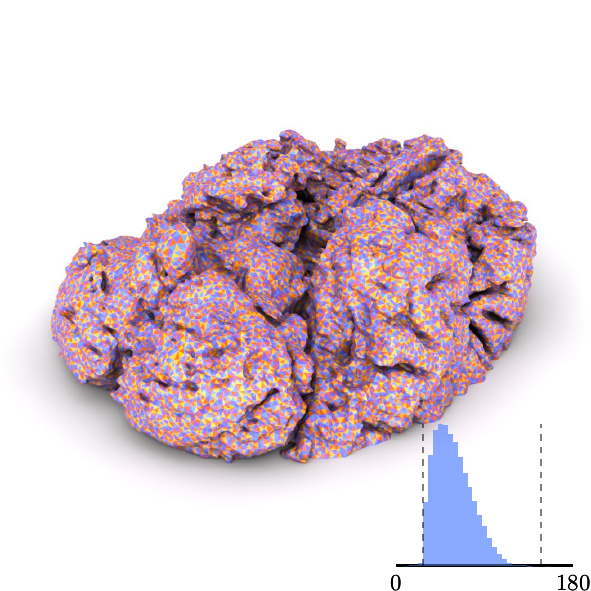}
      \caption{$30^\circ < \theta < 120^\circ$, $cl_{\text{min}} = 1$ (\#f=86,734)}
  \end{subfigure}%
  \hfill%
  \hfill%

  \caption{Intrinsic meshing of three models from the Thingi10K data set (ID: 37278, 39878 and 44374) for different constraints on the intrinsic angles and minimizing the number of elements.}
  \label{fig:examples}
\end{figure*}

\subsection{Robustness}

As previously discussed, all results were computed using an implementation of the MMP algorithm provided by one of the authors of \cite{surazhsky2005fast}. The primary modification to this implementation consists in adapting the minimum distance computation to incorporate an A* search variant of the algorithm. In addition, the structure of the output points was revised: instead of returning three-dimensional coordinates, the algorithm now produces parametric coordinates on the surface, which offer a more robust representation. Finally, the numerical tolerances were adjusted to better reflect the limitations of floating-point arithmetic.

To evaluate the robustness of our method, we conducted a large-scale study using the Thingi10k dataset \cite{zhou2016thingi10k}. We employed the dataset's filtering tools to extract all models that satisfy three strict criteria required by the method: the surfaces must be manifold, closed and consistently oriented. Indeed, the orientation of triangles is used to correctly compute intrinsic angles. The filtering process removes meshes with edges that are not adjacent to two triangles with the same orientation. However, the challenging 'real-world' artifacts such as high-aspect-ratio/degenerate triangles, varying sampling densities and complex topologies that are characteristic of the Thingi10k collection are not filtered out. The resulting test suite provides a rigorous benchmark, presented in Table~\ref{tab:thingi10k_results} for the robustness of our implementation on topologically sound but geometrically complex surfaces.

The largest possible element size and angular bounds between $20$ and $140$ degrees are enforced as constraints throughout the process. It should be noted that the computation of circumcenters may occasionally fail due to numerical inaccuracies. However, such failures can be safely ignored without significantly affecting the overall algorithmic outcome. Consequently, these cases are not classified as failures. In contrast, a run is considered a failure only if a geodesic cannot be computed or if the limits of 2 hours of compute time or 10 GB of memory are exceeded.

Because circumcenters are not always available, the prescribed angle constraints cannot be systematically enforced. Moreover, certain geometries intrinsically prevent the satisfaction of these constraints due to their underlying shape characteristics. As a result, among all successful test cases, only $27.5\%$ strictly satisfy the prescribed angular bounds, while the remaining cases exhibit one or more angles outside the specified range. Despite this, all successful cases converge within fewer than 25 iterations of the main loop.

\begin{table}
\caption{Summary of results on the filtered Thingi10k dataset for intrinsic remeshing with the largest elements and angles bounded between $20^\circ$ and $140^\circ$.}
\label{tab:thingi10k_results}
  \begin{tabular}{lcc}
    \toprule
    Outcome               & \#models  & \%models  \\
    \midrule
    Success               & 4943      & 99.6     \\
    Geodesic error        & 8         & 0.16      \\
    Time limit            & 6         & 0.12      \\
    Memory limit          & 3         & 0.06      \\
    Not manifold vertices          & 3         & 0.06      \\
  \bottomrule
\end{tabular}
\end{table}

\subsection{Performance}

\begin{figure}[t]
  \centering
  \includegraphics{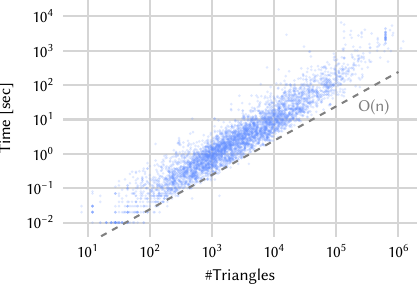}
  \caption{Compute time with respect to model size.}
  \label{fig:timings}
\end{figure}

All computations were carried out on a Genoa (3.25 GHz) CPU using a single-threaded implementation. The corresponding computation times are presented in Fig.~\ref{fig:timings}. The two main computational bottlenecks of the method are the evaluation of geodesic paths and the computation of circumcenters. Even if these constructs are stored to prevent redundant calculations, they remain the dominant contributors to the overall computational cost, primarily due to the expense of evaluating geodesic distances on the surface.

The performance of the meshing method could thus be substantially enhanced through multithreading or GPU acceleration. A straightforward approach would be to use parallel implementations of continuous Dijkstra algorithms. However, due to the inherently sequential nature of Dijkstra’s ordered propagation, parallelization will only achieve limited speed-up.

A more ambitious strategy would involve performing intrinsic local optimization in parallel. This approach, however, necessitates a more sophisticated implementation to ensure that local optimizations do not interfere with other regions of the mesh being updated simultaneously. In \cite{marot2019one}, such an implementation was shown to be efficient for large-scale standard meshing methods, while also highlighting the inherent challenges of parallelizing such computations.

\subsection{Direct High-Order Meshing}

\begin{figure}[htbp]
  \centering
  \hfill%
  \begin{subfigure}{.4\linewidth}
  \centering
    \includegraphics[width=\linewidth]{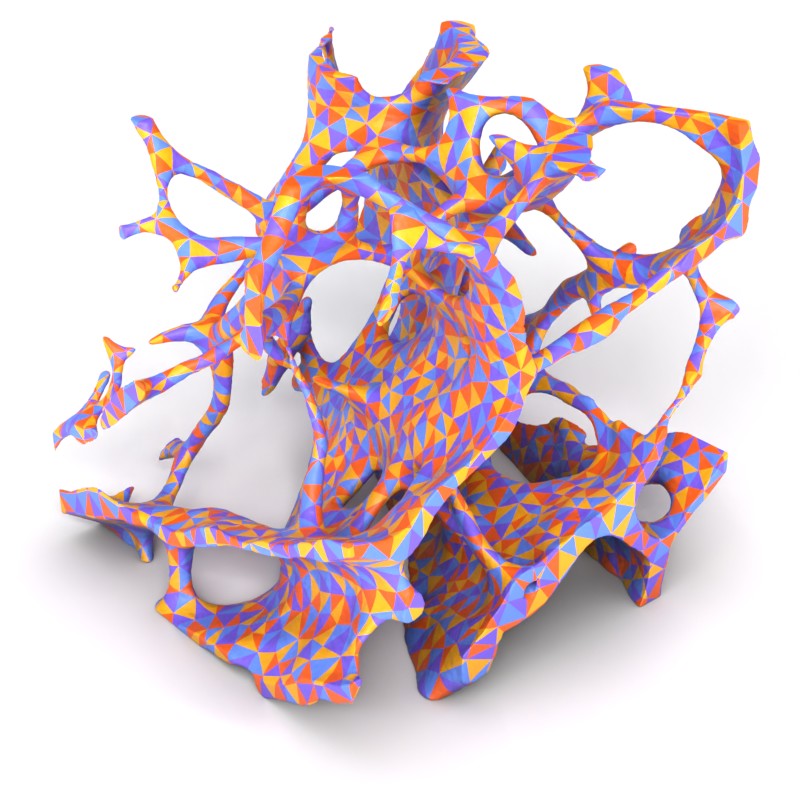}
  \end{subfigure}%
  \hfill%
  \begin{subfigure}{.4\linewidth}
  \centering
    \includegraphics[width=\linewidth]{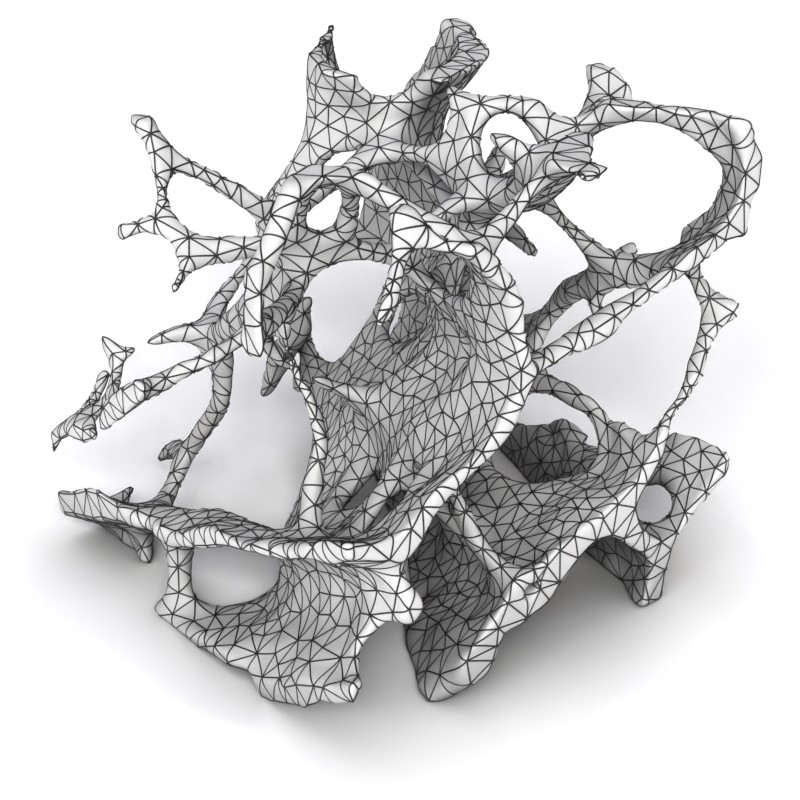}
  \end{subfigure}
  \hfill%
  \hfill%
  \caption{Intrinsic remeshing of a trabecular bone and a third-order polynomial fit on this mesh.}
  \label{fig:bone}
\end{figure}

One important application of intrinsic meshes is the fitting of high-order polynomial elements. This is of significant interest because it provides a direct approach for constructing high-order elements by taking the true geometry and curvature into account from the beginning. This contrasts with classical high-order meshing methods, where linear simplices are first generated and then curved through an optimization procedure to obtain high-order elements \cite{toulorge2013robust}. An example is shown in Fig.~\ref{fig:bone}, where the surface of a trabecular bone is first meshed using intrinsic meshing, and the resulting triangles are then fitted with high-order polynomial elements to produce a high-order mesh composed of classical elements suitable for simulation with standard methods.

It should be noted that these results are obtained solely by constraining edge lengths and angles. In Sec.~\ref{sec:quality}, the notion of triangle quality was introduced to recover a measure of element quality suitable for numerical simulations. Indeed, triangle angles are commonly used as indicators of simulation accuracy in classical numerical methods, although this assumption may not necessarily hold for intrinsic triangulations. Consequently, depending on the intended application of the intrinsic mesh, it may be advantageous to introduce additional quality measures for both triangles and edges. More generally, the method described here is flexible and can accommodate alternative edge or triangle quality criteria.

%% file: future.tex
\section{Perspectives}

In this work, we presented a complete method for intrinsic remeshing, from the technical computation of shortest-path geodesics and intrinsic circumcenters, to the development of a remeshing algorithm that leverages local optimization operations to enforce various constraints on the resulting mesh. More sophisticated operations could be considered, such as locally repositioning vertices within their cavities to further improve mesh quality. Such strategies could also enable more advanced remeshing approaches, including optimal Delaunay triangulation~\cite{sharp2019navigating}.

It was also demonstrated that, once an intrinsic triangulation is obtained, it can readily serve as a basis for fitting different types of elements, such as high-order polynomials. Of particular interest, however, is the potential to use these intrinsic meshes directly, without additional processing. Since the mesh preserves the exact geometry, computations performed on it can fully leverage this accuracy.

An important direction for future work is the extension of the proposed method to open surfaces, as its current restriction to closed surfaces constitutes a significant limitation. In addition, the method does not adequately capture sharp features, which may be interpreted as embedded edges on the surface. Consequently, enabling the representation of boundaries and embedded feature edges remains a key challenge. This difficulty stems from the fact that such edges do not necessarily coincide with shortest paths between vertices and therefore may not correspond to geodesics on the surface. As a result, intrinsic remeshing on open surfaces, as well as the treatment of embedded edges, will require specialized handling of these features.

%% file: appendix.tex
\section{Circumcenter Computation With Pseudo-Sources}\label{sec:circumcenterformula}

As detailed in Sec.~\ref{sec:circumcenter}, it is necessary to compute the position of the circumcenter, if it exists, defined by three pseudo-sources lying in the same plane.
Each pseudo-source is characterized by its position in the plane and its distance to an associated source.
The objective is to determine the coordinates of a point that is equidistant from the three associated sources.
This problem reduces to solving the following system of equations for the unknowns
$x$ and $y$, given three pseudo-sources $A$, $B$, and $C$:
\begin{equation}
\begin{aligned}
  &d_A + \sqrt{(x - x_A)^2 + (y - y_A)^2}
  \\
  =
  &d_B + \sqrt{(x - x_B)^2 + (y - y_B)^2}
  \\
  =
  &d_C + \sqrt{(x - x_C)^2 + (y - y_C)^2}
\end{aligned}
\end{equation}

A first simplification of the system consists in applying a translation of the coordinate system relative to the first pseudo-source. Additionally, if the pseudo-sources are ordered by increasing distance to their associated sources, the problem can be reformulated in terms of a shifted total distance:
\begin{equation}
\begin{gathered}
  \begin{cases}
    \sqrt{{x'}^2 + {y'}^2}
    = \,
    d_B + \sqrt{(x' - x'_B)^2 + (y' - y'_B)^2}
    \\
    \sqrt{{x'}^2 + {y'}^2}
    = \,
    d_C + \sqrt{(x' - x'_C)^2 + (y' - y'_C)^2}
  \end{cases}
\end{gathered}
\end{equation}
  with $d' = d - d_A$, $x' = x - x_A$ and $y' = y - y_A$.
A rotation of the coordinate system eliminates an additional parameter. The final step in this simplification is to scale the system by the inverse of $x'_B$, thereby removing this parameter from the formulation:
\begin{equation}
    \begin{cases}
        X = (x'_B x' + y'_B y') \big/ (\sqrt{{x'_B}^2 + {y'_B}^2} x'_B)
        \\
        Y = (- y'_B x' + x'_B y') \big/ (\sqrt{{x'_B}^2 + {y'_B}^2} x'_B)
    \end{cases}
\end{equation}
In summary, the problem can be expressed in the following simplified form:
\begin{equation}
\begin{gathered}
    \begin{cases}
        \sqrt{X^2 + Y^2}
        =
        D_B + \sqrt{(X - 1)^2 + Y^2}
        \\
        \sqrt{X^2 + Y^2}
        =
        D_C + \sqrt{(X - X_C)^2 + (Y - Y_C)^2}
    \end{cases}
    \\
    \begin{aligned}
        \text{with} \quad
        D_B = d'_B / x'_B \qquad D_C = d'_C / x'_B \\
        X_C = (x'_B x'_C + y'_B y'_C) \big/ (\sqrt{{x'_B}^2 + {y'_B}^2} x'_B) \\
        Y_C = (x'_B y'_C - y'_B x'_C) \big/ (\sqrt{{x'_B}^2 + {y'_B}^2} x'_B) \\
    \end{aligned}
\end{gathered}
\end{equation}

Assuming that the parameters $D_B$ and $D_C$ are nonzero, the resulting system of two equations with two unknowns describes the intersection of two hyperbolas, each having a pseudo-source as a focus.
Furthermore, since both $D_B$ and $D_C$ are positive, the solution is confined to the intersection of specific branches of these hyperbolas.
As a result, two distinct solutions are expected.
By isolating the last term in each equation and then squaring, a linear relationship between $X$ and $Y$ can be established.
Crucially, this squaring does not introduce spurious solutions, as all terms involved are nonnegative:
\begin{equation}
\begin{aligned}
  2 D_B Y_C Y
  &=
  2 (D_C - D_B X_C) X
  \\
  &+ \big( D_C (D_B^2 - 1) + D_B (X_C^2 + Y_C^2 - D_C^2) \big)
\end{aligned}
\end{equation}
Assuming that $Y_C$ is nonzero, $Y$ can be substituted from the derived relation into the first equation.
This reduces the system to a quadratic equation in $X$, which yields the two expected solutions.
The problem can be solved using symbolic resolution, and the resulting solutions are presented in Table~\ref{tab:circumcenter}.

The above reasoning did not consider the cases where $D_B$, $D_C$, or $Y_C$ are equal to zero.
These scenarios correspond, respectively, to the intersection of one line and one hyperbola, the intersection of two lines and the intersection of two hyperbolas whose foci are collinear.
Nonetheless, it remains possible to apply a similar approach to determine the solutions for these special cases.
All valid solutions are presented in Table~\ref{tab:circumcenter}.

\begin{table*}
  \caption{Solutions for the circumcenter position under various parameter conditions.}
  \label{tab:circumcenter}
    \centering
    \renewcommand{\arraystretch}{1.5}
    \begin{tabular}{ll}
        \toprule
        Two Lines ($D_C = D_B = 0$) \\
        \midrule
        $ X = 1 / 2 $ \\
        $ Y = (X_C^2 + Y_C^2 - X_C) / (2 Y_C) $ \\
        \addlinespace \\
        Collinear Points ($Y_C = 0$) \\
        \midrule
        $ X = \big( D_C (D_B^2 - 1) - D_B (D_C^2 - X_C^2) \big) / C $ \\
        $ Y = B / C \ \text{or} \ - B / C $ \\
        $ B = - (D_B^2 - 1) (D_C^2 - X_C^2) \big( (D_B - D_C)^2 - (X_C - 1)^2 \big) $ \\
        $ C = 2 (D_B X_C - D_C) $ \\
        \addlinespace \\
        One Line \& One Hyperbola ($D_B = 0$) \\
        \midrule
        $ X = 1 / 2$ \\
        $ Y = (A + B) / C \ \text{or} \ (A - B) / C$ \\
        $ A = Y_C (X_C - (X_C^2 + Y_C^2 - D_C^2)) $ \\
        $ B = D_C \sqrt{(X_C^2 + Y_C^2 - D_C^2) (X_C^2 + Y_C^2 - D_C^2 + 1 - 2 X_C)} $ \\
        $ C = - 2 (Y_C^2 - D_C^2) $ \\
        \addlinespace \\
        One Hyperbola \& One Line/Hyperbola \\
        \midrule
        $ X = (A + B) / C \ \text{or} \ (A - B) / C$ \\
        $ Y = \left( 2 (D_C - D_B X_C) X + \big( D_C (D_B^2 - 1) + D_B (X_C^2 + Y_C^2 - D_C^2) \big) \right) \big/ (2 D_B Y_C) $ \\
        $ A = X_C (D_B D_C - X_C) (D_B^2 - 1) + (X_C^2 + Y_C^2 - D_C^2) \big( D_B^2 - 1 + D_B (D_B X_C - D_C) \big) $ \\
        $ B = D_B Y_C \sqrt{(D_B^2 - 1) (X_C^2 + Y_C^2 - D_C^2) \big( D_B^2 - 1 - 2 (D_B D_C - X_C) - (X_C^2 + Y_C^2 - D_C^2) \big)} $ \\
        $ C = 2 (D_B^2 - 1) (X_C^2 + Y_C^2 - D_C^2) + 2 (D_B D_C - X_C)^2 $ \\
        \bottomrule
    \end{tabular}
\end{table*}

\newpage

\section{Additional pseudo-codes}

\begin{algorithm}[H]
  \caption{\textsc{symbolicSwaps}($E, allowInsert$)}
  \label{alg:symbolicSwap}
  \begin{algorithmic}[1]
    \STATE Create a symbolic copy $C$ of the mesh
    \STATE Initialize a set $S$ containing all edges $E$
    \WHILE{$S \neq \emptyset$}
      \STATE Extract an edge $e$ from $S$
      \IF{\NOT \textsc{doWeSwap}($e$)}
        \STATE \textbf{continue}
      \ENDIF
      \IF{\textsc{canWeSwap}($e$)}
        \STATE Swap edge $e$ in $C$
        \STATE $S \leftarrow S \cup \{\text{edges adjacent to } e\}$
        \STATE \textbf{continue}
      \ENDIF
      \IF{\NOT $allowInsert$}
        \RETURN \FALSE
      \ENDIF
      \STATE Compute the midpoint $p$ of edge $e$
      \STATE Construct the sets of new and border edges induced by inserting $p$: $E_{\text{new}}$ and $E_{\text{border}}$
      \IF{\textsc{checkIntersections}($E_{\text{new}}, E_{\text{border}}$)}
        \RETURN \FALSE
      \ENDIF
      \STATE Split edge $e$ at $p$ in $C$
      \STATE $S \leftarrow S \cup \{\text{edges adjacent to } e\}$
    \ENDWHILE
    \RETURN \TRUE
  \end{algorithmic}
\end{algorithm}

\begin{algorithm}[H]
  \caption{\textsc{checkIntersections}($E_{\text{new}}, \\E_{\text{border}}$)}
  \label{alg:checkIntersections}
  \begin{algorithmic}[1]
    \FORALL{new edge $n$ in $E_{\text{new}}$}
      \FORALL{border edge $b$ in $E_{\text{border}}$}
          \IF{$n$ and $b$ do intersect}
              \RETURN \TRUE
          \ENDIF
      \ENDFOR
      \FORALL{new edge $m$ in $E_{\text{new}}$}
          \IF{$n\neq m$ \AND $n$ and $m$ do intersect}
              \RETURN \TRUE
          \ENDIF
      \ENDFOR
    \ENDFOR
    \RETURN \FALSE
  \end{algorithmic}
\end{algorithm}

\begin{algorithm}[H]
  \caption{\textsc{TrySplitEdge}($e$)}
  \label{alg:trySplitEdge}
  \begin{algorithmic}[1]
    \STATE Compute the midpoint $m$ of edge $e$
    
    \STATE Construct the sets of new and border edges induced by inserting $m$ in $e$: $E_{\text{new}}$ and $E_{\text{border}}$
    \IF{\textsc{checkIntersections}($E_{\text{new}}, E_{\text{border}}$)}
      \RETURN \FALSE
    \ENDIF

    \STATE $E_{\text{adj}} \gets E_{\text{new}} \cup E_{\text{border}}$
    \IF{\NOT \textsc{symbolicSwaps}($E_{\text{adj}}$, true)}
      \RETURN \FALSE
    \ENDIF

    \STATE $q_{\text{created}} \gets$ the minimum quality among created triangles
    \STATE $q_{\text{removed}} \gets$ the minimum quality among removed triangles
    \IF{$q_{\text{created}} < 0$ \AND $q_{\text{created}} < q_{\text{removed}}$}
      \RETURN \FALSE
    \ENDIF
    
    \STATE Split edge $e$ at $m$ and perform edge swaps
    \RETURN \TRUE
  \end{algorithmic}
\end{algorithm}

\begin{algorithm}[H]
  \caption{\textsc{TryCollapseEdge}($e$)}
  \label{alg:tryCollapseEdge}
  \begin{algorithmic}[1]

    \IF{\NOT \textsc{InDiskCavity}($e$)}
      \RETURN \FALSE
    \ENDIF

    \STATE Let $v_1, v_2, v_3$ be the two endpoints and midpoint of $e$
    \STATE Initialize $q_{\text{created}}^v \leftarrow -\infty$ for all $v \in \{v_1, v_2, v_3\}$
    \STATE Initialize $q_{\text{removed}}^v \leftarrow 0$ for all $v \in \{v_1, v_2, v_3\}$

    \FORALL{$v \in \{v_1, v_2, v_3\}$}
      \STATE Construct the sets of new and border edges induced by collapsing $e$ onto $v$: $E_{\text{new}}$ and $E_{\text{border}}$

      \IF{\textsc{CheckIntersections}($E_{\text{new}},\, E_{\text{border}}$)}
        \STATE \textbf{continue}
      \ENDIF

      \STATE $E_{\text{adj}} \leftarrow E_{\text{new}} \cup E_{\text{border}}$

      \IF{\NOT \textsc{SymbolicSwaps}($E_{\text{adj}}$, false)}
        \STATE \textbf{continue}
      \ENDIF

      \STATE $q_{\text{created}}^{v} \gets$ the minimum quality among created triangles
      \STATE $q_{\text{removed}}^{v} \gets$ the minimum quality among removed triangles

    \ENDFOR

    \STATE $v^{\star} \leftarrow \arg\max_{v}\, (q_{\text{created}}^{v} - q_{\text{removed}}^{v})$

    \IF{$q_{\text{created}}^{v} < 0$ \AND $q_{\text{created}}^{v} < q_{\text{removed}}^{v}$}
      \RETURN \FALSE
    \ENDIF

    \STATE Collapse $e$ onto $v^{\star}$ and perform edge swaps
    \RETURN \TRUE
  \end{algorithmic}
\end{algorithm}

\begin{algorithm}[H]
  \caption{\textsc{TrySplitTriangle}($t$)}
  \label{alg:trySplitTriangle}
  \begin{algorithmic}[1]
    \STATE Compute the circumcenter $c$ of the triangle $t$
    \IF {$c$ is not found}
        \RETURN \FALSE
    \ENDIF

    \STATE Find the triangle $t'$ that contains the point $c$.
    \STATE Construct the sets of new and border edges induced by inserting $c$ in $t'$: $E_{\text{new}}$ and $E_{\text{border}}$    
    \IF{\textsc{checkIntersections}($E_{\text{new}}, E_{\text{border}}$)}
      \RETURN \FALSE
    \ENDIF

    \STATE $E_{\text{adj}} \gets E_{\text{new}} \cup E_{\text{border}}$
    \IF{\NOT \textsc{symbolicSwaps}($E_{\text{adj}}$, true)}
      \RETURN \FALSE
    \ENDIF

    \STATE $q_{\text{created}} \gets$ the minimum quality among created triangles
    \STATE $q_{\text{removed}} \gets$ the minimum quality among removed triangles
    \IF{$q_{\text{created}} < 0$ \AND $q_{\text{created}} < q_{\text{removed}}$}
      \RETURN \FALSE
    \ENDIF
    
    \STATE Split triangle $t'$ at $c$ and perform edge swaps
    \RETURN \TRUE
  \end{algorithmic}
\end{algorithm}